\documentclass[a4paper,11pt]{article}
\pdfoutput=1

\usepackage{jcappub}
\usepackage[T1]{fontenc}
\usepackage{booktabs}
\usepackage{siunitx}
\usepackage{lineno}
\usepackage{placeins}
\usepackage{xcolor}

\graphicspath{{figures/}}

\title{\boldmath Emulating the nonlinear effects of modified gravity on the matter power spectrum for reconstruction}

\author[a,b,c]{Lanyang Yi,}
\author[c,d,e]{Kazuya Koyama,}
\author[a]{Qi Xiong}
\author[a,b,f]{and Gong-Bo Zhao}

\affiliation[a]{National Astronomical Observatories, Chinese Academy of Sciences,\\
Beijing 100101, P.R.China}

\affiliation[b]{School of Astronomy and Space Sciences, University of Chinese Academy of Sciences,\\
Beijing 100049, P.R.China}

\affiliation[c]{Institute of Cosmology and Gravitation, University of Portsmouth,\\
Dennis Sciama Building, Burnaby Road, Portsmouth PO1 3FX, United Kingdom}

\affiliation[d]{Kavli Institute for the Physics and Mathematics of the Universe (WPI), UTIAS,\\
The University of Tokyo, Kashiwa, Chiba 277-8583, Japan}

\affiliation[e]{Yukawa Institute for Theoretical Physics, Kyoto University,\\
Kitashirakawa Oiwakecho, Sakyo-ku, Kyoto 606-8502, Japan}

\affiliation[f]{Institute for Frontiers in Astronomy and Astrophysics, Beijing Normal University,\\
Beijing 102206, P.R.China}

\emailAdd{lanyang.yi@port.ac.uk}
\emailAdd{kazuya.koyama@port.ac.uk}
\emailAdd{xiongqi@bao.ac.cn}
\emailAdd{gbzhao@nao.cas.cn}

\abstract{Including nonlinear information from modified gravity (MG) brings more constraining power to the model-independent reconstruction of MG functions. However, the calculation of the nonlinear matter power spectrum with \texttt{MGCAMB}+\texttt{ReACT} is expensive in repeated likelihood evaluations, which limits the exploration of the high-dimensional parameter space. In this work, we construct a neural-network emulator for the nonlinear correction $R_{\mathrm{MG}}(k,z)=P_{\mathrm{MG}}^{\mathrm{NL}}(k,z)/P_{\mathrm{MG}}^{\mathrm{L}}(k,z)$, which is trained with \texttt{CosmoPower} using 209772 successfully computed spectra and validated with representative spectra, an independent validation set, MCMC tests with synthetic data and an out-of-range test. For $\Lambda$CDM and moderate MG nonlinear corrections, the emulated power spectra agree with the reference predictions to within $1.5\%$ over the full scale range considered. For the extreme nonlinear case, the same accuracy is retained for $k<1\,\mathrm{Mpc}^{-1}$. Over the independent validation set, the mean residual is close to zero, with the $2\sigma$ scatter below $1\%$ for $k<0.5\,\mathrm{Mpc}^{-1}$ and about $2\%$ on smaller scales. The synthetic data MCMC analyses recover the input $\Lambda$CDM cosmology and the GR limits of the reconstructed MG functions within the posterior uncertainties, supporting the consistency of the emulator results for the synthetic-data configurations considered. The test with out-of-range points further supports this conclusion with the $2\sigma$ region extending up to approximately $4\%$. We also demonstrate three applications: using $R_{\mathrm{MG}}$ to distinguish models with nearly degenerate linear power spectra, using the emulator for theory predictions for current photometric-survey $3\times2$pt likelihoods, and forecasting DESI+CSST constraints with principal component analysis.
}


\compress

\begin{document}

\maketitle

\section{Introduction}

Since the discovery of cosmic acceleration through observations of the redshift-distance relation of Type Ia supernovae~\cite{Riess:1998,Perlmutter:1999}, understanding the nature of this mechanism has become one of the primary questions in cosmology. A widely accepted explanation is that the acceleration is driven by a new component of energy density, called dark energy, with negative pressure $p$. One simple and widely accepted model is the cosmological constant $\Lambda$. However, recent DESI results~\cite{DESI:DR1FS,DESI:DR2BAO} in combination with CMB and three supernovae datasets~\cite{DES:dovekie,union3.1:pantheon+} have shown evidence in favour of dynamical dark energy over $\Lambda$ at the level $3.4\sigma$ (DES Dovekie), $3.4\sigma$ (Union3.1) and $3.2\sigma$ (New Pantheon+), which agree well and motivate exploring alternative models to fit the observations. One possibility is that the acceleration arises from modifications to General Relativity (GR) on cosmological scales, often referred to as modified gravity (MG)~\cite{Clifton:2011jh,Koyama:2015vza,Ishak:2018his}. Distinguishing between these theories requires precise measurements of the expansion history and the growth of cosmic structure from current and future surveys.

Following the conventions of Ref.~\cite{Ma:NewtonGauge}, we start our modification to GR from the perturbed Friedmann-Robertson-Walker metric in the Newtonian gauge:
\begin{equation}
    ds^2 = -(1+2\psi)dt^2 + a^2(t)(1-2\phi)d\vec{x}^2,
\end{equation}
where $\psi$ and $\phi$ are the Newtonian potential and the spatial curvature perturbation, respectively. In GR, these two potentials are equal when anisotropic stress is neglected, and their relation to matter perturbations is given by the Poisson equation and the Weyl-potential relation: 
\begin{align}
    k^2 \psi &=
    -4\pi G\,a^2 \rho \Delta, \\
    k^2 \frac{\phi+\psi}{2} &=
    -4\pi G\,a^2 \rho \Delta,
\end{align}
where $G$ is the Newton's gravitational constant, $k$ is the Fourier wavenumber, $\rho$ is the background density of matter and $\Delta$ is the comoving matter density contrast. In MG theories, these relations can be modified, leading to different predictions for the growth history of structure. To capture these modifications without specifying a particular MG theory, we can introduce two functions $\mu(k,a)$ and $\Sigma(k,a)$ that modify these two equations. These two functions are widely used in phenomenological tests of gravity on cosmological scales
~\cite{Caldwell:2007cw,Amendola:2007rr,Hu:2007pj,Jain:2007yk,Bertschinger:2008zb,Zhao:2008bn,Pogosian:2010tj,Zhao:2010dz,Silvestri:2013ne, Pogosian:2021mio} and can be constrained by various cosmological observations, such as redshift-space distortions (RSD)~\cite{DESI:DR1CPE,DESI:DR1MG}, weak lensing (WL)~\cite{DES:Y3extend}, and cosmic microwave background (CMB)~\cite{Planck:2018Planck,WAS:MG}. The most popular way to constrain these functions in recent galaxy survey results is to assume a specific model, e.g. DES parameterization~\cite{Ferreira:2010sz,DES:Y1extend}. However, testing gravity in a model-dependent way may miss some important features of the theory and lead to biased results. Meanwhile, many recent works constrain perturbations assuming a $\Lambda$CDM background~\cite{DESI:DR1CPE,DES:Y3extend,Planck:2018Planck,WAS:MG}, or simply assume the background expansion history to be $w_0 w_a$CDM~\cite{DESI:DR1MG}. Considering the fact that the expansion history can affect the growth of structure, it is worth testing the background expansion together with perturbations at the same time in a model-independent way, going beyond the analysis of specific models. Therefore, a model-independent reconstruction of these functions is important for testing gravity on cosmological scales.

On the other hand, large-scale-structure data from spectroscopic and photometric surveys also contain a non-negligible amount of information on small scales, where nonlinear effects significantly affect the structure formation. In previous gravity tests, one often cuts the nonlinear regime and uses only the information from linear growth~\cite{DES:Y1extend,Zucca:2019MGCAMB,DES:Y3extend}, especially for the weak lensing data. However, MG theories can also affect the nonlinear evolution of structure~\cite{Koyama:2009me}, which is crucial for interpreting data from spectroscopic and photometric surveys. Including this nonlinear information to fit with MG predictions can significantly improve the constraints on MG theories~\cite{Wang:2024mge}. Therefore, accurate predictions of the nonlinear matter power spectrum in MG theories are essential for testing these models against observations. Using a Boltzmann code together with a halo model that characterises MG nonlinear effects, e.g. the \textit{halo model reaction} (HMR)~\cite{Cataneo:2018cic,Bose:2020,Bose:2023}, we can achieve this goal. Through this method, however, the nonlinear spectrum calculation is very time-consuming, making it difficult to use in a reconstruction pipeline where we need to evaluate the spectrum with many more parameters than simple MG models. One possible solution is to use an emulator, which is a machine learning model that can predict the nonlinear correction to the matter power spectrum for a given MG model, to speed up the calculation. This allows us to explore the parameter space of MG reconstruction and compare theory predictions with observations faster and more efficiently.

This paper focuses on emulating the nonlinear effects of MG on the matter power spectrum in the reconstruction pipeline. Simulation-based emulators have been developed for several MG and other beyond--$\Lambda$CDM theories~\cite{Winther:2019mus,Ramachandra:2020lue,Arnold:2021xtm,Saez-Casares:2023olw,Mauland:2023pjt,Fiorini:2023fjl,Srinivasan:2026fxo}. An alternative strategy is to emulate theory predictions generated by an analytical code~\cite{Tsedrik:2024cdi}. To accelerate the theory calculation, we adopt the latter strategy and train our emulator on spectra computed with \texttt{MGCAMB}+\texttt{ReACT}. We will validate the emulator with several methods, including specific-model tests, validation-set tests, and MCMC tests using synthetic data, to show its accuracy and reliability on both cosmological and MG parameters. We also show some applications of the emulator in MG reconstruction, demonstrating how it can support studies of nonlinear MG effects.

The paper is organized as follows. Section~\ref{sec:background} introduces the theoretical background of modified gravity reconstruction and nonlinear corrections. Section~\ref{sec:emulator-training} describes the emulator training method. Section~\ref{sec:validation-method} presents the emulator validation strategy. Section~\ref{sec:validation-results} reports the validation results, while Section~\ref{sec:application} shows some applications of the emulator. We summarize our conclusions in Section~\ref{sec:discussion}.

\section{Theoretical Background}
\label{sec:background}

\subsection{Modified Gravity Reconstruction}

Based on scalar-tensor theories, the MG reconstruction is a useful approach to investigate the deviations from GR and the evolution of gravity in a model-independent way. Our reconstruction method follows the framework developed in Ref.~\cite{Pogosian:2021mio}. The late-time background evolution is written in terms of an effective dark-energy density function $\Omega_X(a)$,
\begin{equation}
    \left(\frac{da}{dt}\right)^2
    =
    a^2 H_0^2
    \left[
        \frac{\Omega_r}{a^4}
        +
        \frac{\Omega_m}{a^3}
        +
        \Omega_X(a)
    \right],
\end{equation}
where $\Omega_X(a)$ is the effective dark-energy density, describing all the contributions in the universe except for matter and radiation. We set $\Omega_X(a=1)=\Omega_\Lambda$ to make $\Omega_m+\Omega_r+\Omega_X(a=1)=1$ today. Since $\Omega_X$ may become negative in MG theories, which can make the equation of state $w(a)$ singular, we choose $\Omega_X(a)$ as the primary reconstructed quantity instead of $w(a)$.

Focusing on scalar perturbations in Newtonian gauge, we use two phenomenological functions in the reconstruction, $\mu$ and $\Sigma$, which are defined as follows:
\begin{align}
    k^2 \psi &=
    -4\pi G\,\mu(a,k)\,a^2 \rho \Delta,
    \label{eq:modified-poisson} \\
    k^2 \frac{\phi+\psi}{2} &=
    -4\pi G\,\Sigma(a,k)\,a^2 \rho \Delta .
    \label{eq:weyl-poisson}
\end{align}
The function $\mu$ is related to the potential $\psi$ which affects the peculiar velocities of matter and therefore controls the clustering of matter, which can be observed by redshift-space distortions (RSD) and the growth of structure. The function $\Sigma$ is directly connected with the Weyl potential $\phi+\psi$, which affects the propagation of light and can be observed by weak lensing and CMB lensing. Their GR limits are $\mu=\Sigma=1$.

Following the implementation of Ref.~\cite{Pogosian:2021mio}, we assume a scale-independent MG theory on linear scales and focus on the redshift evolution of these functions. This is motivated by the fact that, in many scalar--tensor theories, the relevant scale dependence is often outside the range probed by large-scale-structure data within the linear perturbation theory~\cite{Wang:2012kj,Joyce:2014kja,Pogosian:2016pwr}.

The three functions $\Omega_X(a)$, $\mu(a)$ and $\Sigma(a)$ are represented by several nodes. The reconstruction uses 11 nodes: ten uniformly distributed over $a\in[1,0.25]$ ($z\in[0,3]$), plus one node at $a=0.2$($z=4$). The nodes are ordered so that $\mu_{11}$ and $\Sigma_{11}$ correspond to the $z=0$ node. Note that $\Omega_{X,11}$ is set to the value of $\Omega_\Lambda$. At higher redshift the functions are driven smoothly toward their $\Lambda$CDM/GR values using nine anchor nodes arranged in a tanh-like pattern up to $z=1000$. A cubic spline is then used to construct continuous functions from these nodes.

\subsection{Nonlinear Modified-Gravity Spectrum}

The nonlinear modification of the MG matter power spectrum is based on the framework of Ref.~\cite{Wang:2024mge}. Following the method, we need to calculate the reaction function defined in Ref.~\cite{Cataneo:2018cic} which is the ratio of the real MG nonlinear spectrum and the pseudo MG spectrum,
\begin{equation}
    \mathcal{R}(k,z) =
    \frac{P_{\mathrm{NL}}(k,z)}
    {P^{\mathrm{pseudo}}(k,z)} ,
\end{equation}
where $P_{\mathrm{NL}}$ is the real nonlinear matter power spectrum and $P^{\mathrm{pseudo}}$ is the pseudo MG nonlinear matter power spectrum calculated under the $\Lambda$CDM nonlinear model while its linear clustering obeys the MG theory.

The MG nonlinear modification is introduced through a phenomenological correction to the Poisson equation. On fully nonlinear scales, we can rewrite the Poisson equation as
\begin{equation}
    k^2\psi_{\mathrm{NL}}(k,a)
    =
    -4\pi G\,[1+\mathcal{F}(k,a)]\,a^2\rho \Delta_{\mathrm{NL}}(k,a),
\end{equation}
where $\mathcal{F}(k,a)$ captures the MG nonlinear effects. Ref.~\cite{Wang:2024mge} follows the parameterization used in \texttt{ReACT}, in which the MG nonlinear effects are phenomenologically described by four parameters $p_1$--$p_4$. The parameter $p_1$ quantifies the screening strength, while $p_2$, $p_3$ and $p_4$ are related to the effects that work in the scale-dependent MG models. Since our reconstruction method assumes the MG functions to be scale-independent, we keep only $p_1$ as the free nonlinear parameter and set $p_2=p_3=p_4=0$ to switch off the associated effects. The limits $p_1\rightarrow -\infty$ and $p_1\rightarrow \infty$ respectively correspond to strong screening, where the nonlinear effects go back to the $\Lambda$CDM prediction, and the unscreened case.

In MG theories, for lensing observables, we need to consider the Weyl potential in cosmic shear angular power spectrum calculations. Following the assumption in Ref.~\cite{Wang:2024mge}, we assume that the relation between the Weyl potential and the matter perturbation is consistent with the linear-scale relations given by Eqs.~\eqref{eq:modified-poisson} and \eqref{eq:weyl-poisson}, allowing the nonlinear effects of $P_{WW}(k)$ and $P_{W\delta}(k)$ to be computed from $P_{\delta\delta}(k)$.

\section{Emulator Training Method}
\label{sec:emulator-training}

\subsection{Emulator Target}

The emulator target is not the nonlinear MG matter power spectrum $P_{\mathrm{MG}}^{\mathrm{NL}}(k,z)$ but rather the ratio that quantifies the nonlinear modification relative to the linear prediction:
\begin{equation}
    R_{\mathrm{MG}}(k,z) =
    \frac{P_{\mathrm{MG}}^{\mathrm{NL}}(k,z)}
    {P_{\mathrm{MG}}^{\mathrm{L}}(k,z)}.
\end{equation}
This choice is motivated by the fact that ratio-based targets isolate the model-dependent modifications and reduce the dynamic range that the network must learn~\cite{Winther:2019mus,Ramachandra:2020lue,Arnold:2021xtm,Saez-Casares:2023olw,Mauland:2023pjt,Fiorini:2023fjl,Srinivasan:2026fxo}. In the MG reconstruction model, the Boltzmann code can predict large-scale MG effects reliably, while MG also has a significant impact on the small scales. By emulating the ratio, we can retain the accurate linear prediction to capture the correct large-scale MG effect and focus the emulator on learning the nonlinear modifications, which are more challenging and expensive for the nonlinear pipeline to model. This approach also helps mitigate potential biases in the emulator while trying to emulate the MG effect at large and small scales at the same time.

\subsection{Training Parameters}

The emulator input parameter vector is
\begin{equation}
    \boldsymbol{\theta} =
    \{
    \log A, n_s, h, \Omega_b h^2, \Omega_c h^2,
    \mu_1,\ldots,\mu_{11},
    \Omega_{X,1},\ldots,\Omega_{X,10},
    z, p_1
    \}.
\end{equation}
Here $\log A$ ($\ln (A_s \times 10^{10})$), $n_s$, $h$, $\Omega_b h^2$ and $\Omega_c h^2$ are the standard cosmological parameters. We do not consider the optical depth $\tau$ due to its limited impact on the matter power spectrum. The parameters $\mu_1,\ldots,\mu_{11}$ are the nodes of the reconstructed MG function $\mu(a)$, while $\Omega_{X,1},\ldots,\Omega_{X,10}$ are the sampled effective dark energy density nodes. The last two parameters are the redshift $z$ and the nonlinear screening parameter $p_1$. The $\Sigma$ nodes are not included because the emulator is working on the matter power spectrum, which is not directly affected by $\Sigma$.

\subsection{Training Data}

The training dataset consists of samples from two complementary sources. The first part is from the reconstruction chains from Ref.~\cite{Pogosian:2021mio}; one quarter of the chain samples are selected by Latin Hypercube Sampling (LHS) as a randomized selection from the chain files, using the MG parameters ($\mu$ and $\Omega_X$) of these samples as the input points of the emulator, while for standard cosmological parameters, we generate random samples within their $3\sigma$ posterior ranges from the chains. For $z$ and $p_1$, we choose the range $0 \leq z \leq 5.0$ and $-2.0 \leq p_1 \leq 2.0$ and generate samples randomly within these ranges. These samples preserve the MG parameter combinations accepted in the previous work, which makes the MG part of these input points more physically motivated. Detailed parameter ranges used for sampling are listed in Table~\ref{tab:training-params}.

\begin{table}[!htbp]
    \centering
    \small
    \setlength{\tabcolsep}{3pt}
    \begin{tabular*}{\textwidth}{@{\extracolsep{\fill}}llllll@{}}
        \toprule
        Parameter & Range & Parameter & Range & Parameter & Range \\
        \midrule
        $\log A$ & $[2.90, 3.15]$ & $\mu_1$ & $[0.74, 1.22]$ & $\Omega_{X,1}$ & $[-2.09, 4.15]$ \\
        $n_s$ & $[0.94, 1.00]$ & $\mu_2$ & $[0.61, 1.35]$ & $\Omega_{X,2}$ & $[-1.10, 3.01]$ \\
        $h$ & $[0.60, 0.78]$ & $\mu_3$ & $[0.54, 1.39]$ & $\Omega_{X,3}$ & $[-1.06, 3.16]$ \\
        $\Omega_b h^2$ & $[0.020, 0.024]$ & $\mu_4$ & $[0.58, 1.55]$ & $\Omega_{X,4}$ & $[-0.78, 2.70]$ \\
        $\Omega_c h^2$ & $[0.11, 0.13]$ & $\mu_5$ & $[0.53, 1.88]$ & $\Omega_{X,5}$ & $[0.02, 1.89]$ \\
        $z$ & $[0, 5.0]$ & $\mu_6$ & $[0.59, 2.13]$ & $\Omega_{X,6}$ & $[0.41, 1.02]$ \\
        $p_1$ & $[-2, 2]$ & $\mu_7$ & $[0.48, 2.24]$ & $\Omega_{X,7}$ & $[0.50, 0.85]$ \\
        & & $\mu_8$ & $[0.44, 2.32]$ & $\Omega_{X,8}$ & $[0.47, 0.76]$ \\
        & & $\mu_9$ & $[0.47, 2.37]$ & $\Omega_{X,9}$ & $[0.62, 0.83]$ \\
        & & $\mu_{10}$ & $[0.36, 2.44]$ & $\Omega_{X,10}$ & $[0.54, 0.76]$ \\
        & & $\mu_{11}$ & $[0.17, 3.09]$ & & \\
        \bottomrule
    \end{tabular*}
    \caption{Ranges of the parameters used for generating the training dataset.}
    \label{tab:training-params}
\end{table}

The second part is a set of random samples with all parameters generated from their $3\sigma$ posterior ranges shown in Table~\ref{tab:training-params}. The number of completely random samples is chosen to be twice the number of the first part. This random component broadens the training dataset beyond the accepted parameter space and improves emulator robustness near the edges of the physical parameter space. The combined training set contains approximately $9\times10^5$ candidate data points for the spectrum calculations described below. Each data point corresponds to a specific input parameter vector $\boldsymbol{\theta}$ and its associated target ratio $R_{\mathrm{MG}}(k,z)$ is calculated from the spectrum generation process described in the next subsection.

\subsection{Spectrum Generation}
\label{sec:Generation}

The matter power spectra used for emulator training are based on the MG reconstruction and nonlinear power spectrum model described in Section~\ref{sec:background}. The linear power spectra are computed with \texttt{MGCAMB}\footnote{\url{https://github.com/sfu-cosmo/MGCAMB}}~\cite{Wang:2023MGCAMB,Zucca:2019MGCAMB,Hojjati:2011MGCAMB,Zhao:2008bn}, which is a modified version of the Boltzmann code \texttt{CAMB}\footnote{\url{https://github.com/cmbant/CAMB}}~\cite{Lewis:2000ah,Lewis:2026mif}. The nonlinear MG power spectra are computed using \texttt{MGCAMB} together with \texttt{ReACT}\footnote{\url{https://github.com/nebblu/ReACT}}~\cite{Cataneo:2018cic,Bose:2020,Bose:2023} as the \texttt{MGCAMB} nonlinear extension, following Ref.~\cite{Wang:2024mge}.

For each input parameter point, \texttt{MGCAMB} calculates the MG linear matter power spectrum and the corresponding pseudo nonlinear power spectrum. Using the input parameter and theory predictions from \texttt{MGCAMB}, \texttt{ReACT} computes the reaction function and passes it to \texttt{MGCAMB}, then the nonlinear \texttt{MGCAMB} extension combines it with the pseudo nonlinear power spectrum to obtain the real MG nonlinear power spectrum $P_{\mathrm{MG}}^{\mathrm{NL}}(k,z)$. We calculate the power spectra in the range $10^{-5} \leq k < 10\,\mathrm{Mpc}^{-1}$ with 420 $k$ modes in total at the redshift $z$ specified by the input point. Of the approximately $9\times10^5$ candidate points, 209772 yielded successfully computed spectra and were retained for emulator training; the remaining candidates were rejected by \texttt{MGCAMB}. The emulator is trained to predict the ratio $R_{\mathrm{MG}}(k,z)$ at these $k$ and parameter points.

\subsection{Emulator Training}

The emulator is trained with \texttt{CosmoPower}\footnote{\url{https://github.com/alessiospuriomancini/cosmopower}}~\cite{SpurioMancini:2021ppz}, using the \texttt{cosmopower\_NN} neural-network model. The emulator output is the ratio $R_{\mathrm{MG}}(k,z)$ evaluated on the 420-mode $k$ grid described above, and the inputs are the corresponding cosmological, MG, redshift, and nonlinear parameters.

We use a fully connected neural network with four hidden layers, each containing 512 nodes,
\begin{equation}
    n_{\mathrm{hidden}} = [512,512,512,512].
\end{equation}
The network predicts the full ratio vector over all $k$ modes for each input parameter point. We reserve 20 percent of the training set for validation during training. The optimization uses a five-stage learning-rate schedule,
\begin{equation}
    [10^{-2},10^{-3},10^{-4},10^{-5},10^{-6}].
\end{equation}
At each learning-rate stage we use a batch size of 1024 and early stopping with a patience of 100 epochs, allowing up to 1000 epochs per stage.

\section{Emulator Validation Method}
\label{sec:validation-method}

The emulator is validated in four steps as follows. The first test compares the nonlinear matter power spectrum predicted by the emulator+\texttt{MGCAMB} (referred to as emulator hereafter) with \texttt{MGCAMB}+\texttt{ReACT} reference results (referred to as \texttt{MGCAMB} hereafter) for representative cosmologies. The second test measures the difference distribution over 2000 validation points. The third test validates the emulator in an MCMC analysis using $\Lambda$CDM synthetic data. We additionally assess the emulator predictions at parameter points outside the training ranges that are sampled by the MCMC chains.

\subsection{Representative Spectra Comparison}
\label{sec:spectra-comparison}

We select a small set of physically interpretable cosmologies to test the emulator, including $\Lambda$CDM and several MG models. The cosmologies we choose are based on the fact that their linear power spectra at $z=0$ do not deviate strongly from the $\Lambda$CDM result. These models are not intended to represent a general MG cosmology. Instead, they are chosen as a linear-degenerate test: their linear matter power spectra are required to be nearly indistinguishable from the $\Lambda$CDM prediction at $z=0$, while their nonlinear correction can still differ because of the different growth history. This provides a non-trivial validation test for the emulator, since the test focuses on whether the emulator can recover the nonlinear MG effects. For each cosmology, we compare $P_{\mathrm{emu}}^{\mathrm{NL}}(k,z)$ predicted by the emulator with the reference \texttt{MGCAMB} result and show the comparison of the spectra and their differences,
$\Delta_P(k,z) = \frac{P_{\mathrm{emu}}^{\mathrm{NL}}(k,z)}{P_{\mathrm{MGCAMB}}^{\mathrm{NL}}(k,z)}- 1$ .
The $\Lambda$CDM case tests the GR limit, while the MG cases test the emulator sensitivity to variations in input parameters, especially $\mu_i$, $\Omega_{X,i}$ and $p_1$.

\subsection{Validation Set Comparison}
\label{sec:set-comparison}

The validation set contains 2000 spectra that are not used for training. These consist of 1000 additional points selected from the previous reconstruction chains and 1000 additional randomly generated points, which are sampled and generated using the same method shown in Section~\ref{sec:Generation}. For each validation point, we compute the spectral differences $\Delta_P(k,z)$ using the same definition as in Section~\ref{sec:spectra-comparison} and show the statistical results as a demonstration of the emulator performance over the parameter space.

\subsection{Synthetic Data MCMC Validation}
\label{sec:mcmc}

This MCMC validation is designed as a pipeline consistency test. Since the synthetic data vectors are generated from a known $\Lambda$CDM fiducial model, the goal is to verify that the likelihood theory prediction pipeline with the emulator nonlinear correction can recover the input cosmology and the GR limits of the reconstructed MG functions. This test therefore checks whether replacing the nonlinear calculation by the emulator introduces a detectable bias in parameter constraints. We generate weak lensing angular power spectrum and BAO+RSD synthetic data in $\Lambda$CDM and perform MCMC analyses~\cite{Lewis:2002ah,Lewis:2013hha} and get best-fit results with the minimize sampler BOBYQA~\cite{powell2009bobyqa,Cartis:2018xum,Cartis:2018jxl} in \texttt{Cobaya}\footnote{\url{https://github.com/CobayaSampler/cobaya}}~\cite{Torrado:2019Cobaya,Torrado:2021Cobaya} using the emulator to predict the nonlinear effects of matter power spectrum in weak lensing likelihood. We assess convergence of the MCMC chains by the Gelman-Rubin criterion, assuming the chains have converged when $R-1 \lesssim 0.1$ for the least converged eigenvalue, and use \texttt{GetDist}\footnote{\url{https://github.com/cmbant/getdist}}~\cite{Lewis:2019xzd} to analyze results. In the following section, we will introduce the details of the MCMC parameter setup and the synthetic data.

\subsubsection{DESI-like BAO+RSD Data}

We generate BAO+RSD synthetic data using the Boltzmann code \texttt{CAMB} in the Planck 2018 result~\cite{Planck:2018Planck} $\Lambda$CDM cosmology in two cases. In the first case, we generate the data points in terms of $D_M/r_d$, $D_H/r_d$ and $f\sigma_8$ at redshifts $z=0.295$, $0.510$, $0.706$, $0.919$, $1.317$ and $1.491$, corresponding to the effective redshifts of BGS, the three LRG bins, ELG and QSO in the DESI DR1 Full-Shape measurements~\cite{DESI:DR1FS}, with an additional DESI DR1 Ly$\alpha$ Forest BAO measurement~\cite{DESI:DR1Lya} of $D_M/r_d$ and $D_H/r_d$ at $z=2.330$. The covariance matrix is also taken from the ShapeFit results of the DESI DR1 Full-Shape measurements. This case is considered the DESI DR1-like data, called DESI-I afterwards.

In the second case, we consider 4 different tracers: BGS, LRG, ELG and QSO, all are DESI observation targets, and assume the galaxy bias evolves as $b(z)=b_0/D(z)$ for each tracer, where $D(z)$ is the linear growth factor. Each tracer is considered in one redshift bin and independent of each other. Following the settings in Ref.~\cite{DESI:forecast}, the BGS is in range of $0.0<z<0.4$ and evolves as $b_{BGS}(z)=1.34/D(z)$, the LRG is in range of $0.4<z<1.1$ and evolves as $b_{LRG}(z)=1.7/D(z)$, the ELG is in range of $1.1<z<1.6$ and evolves as $b_{ELG}(z)=0.84/D(z)$, and the QSO is in range of $1.6<z<2.1$ and evolves as $b_{QSO}(z)=1.2/D(z)$. Data points are generated in terms of $D_M/r_d$, $D_H/r_d$ and $f\sigma_8$ at redshifts $z=0.2$, $0.75$, $1.35$ and $1.85$. This case is considered the full DESI-like data, which is a forecast for the full DESI survey results, called DESI-V afterwards.

To generate the covariance in the second case, we calculate the Fisher matrix of three measurements for the full DESI survey, a 14000 $\mathrm{deg}^2$ sky area, using the \texttt{desilike}\footnote{\url{https://github.com/cosmodesi/desilike}}, which is a cosmological inference framework to analyze the DESI data, calculate the likelihood and do the Fisher forecasts for BAO and RSD measurements. For details of the forecast method, see Ref.~\cite{DESI:forecast}. For each tracer, the Fisher matrix is defined as~\cite{White:Fisher}
\begin{equation}
    F_{ij}
    =
    \sum
    \int \frac{V_0\,d^3k}{(2\pi)^3}
    \left(\frac{\partial P}{\partial p_i}\right)
    C^{-1}
    \left(\frac{\partial P}{\partial p_j}\right),
\end{equation}
where $p_i$ and $p_j$ are the measurements we want to forecast, $P$ is the galaxy power spectrum, $C$ is the covariance matrix of the power spectrum, and $V_0$ is the survey volume. To calculate it and forecast the measurements we need, we follow the Fisher-forecast prescription in Section 6.1 of Ref.~\cite{DESI:forecast}, describe the large-scale RSD signal using the linear Kaiser spectrum multiplied by a phenomenological damping factor,
\begin{equation}
    P(k,\mu,z) = \left(b(z)+f(z)\mu^2\right)^2 P_m(k,z) A(k,\mu,z),
\end{equation}
where $P_m(k,z)$ is the linear matter power spectrum, $\mu$ is the cosine of the angle between the wave vector $\mathbf{k}$ and the line of sight, and $A(k,\mu,z)$ is the BAO damping factor, which is given by
\begin{equation}
    A(k,\mu,z) = \exp\left[-\frac{k^2}{2}\left(\mu^2\Sigma_\parallel^2(z) + (1-\mu^2)\Sigma_\perp^2(z)\right)\right].
\end{equation}
Here the $\Sigma_\parallel(z)$ and $\Sigma_\perp(z)$ are Lagrangian displacement distances, given by
\begin{equation}
     \Sigma_\perp(z) = 9.4\left(\frac{\sigma_8(z)}{0.9}\right)h^{-1}\mathrm{Mpc},\quad \Sigma_\parallel(z) = (1+f(z))\Sigma_\perp(z),
\end{equation}
and are both multiplied by a factor ($\in$ [0.5, 1]) to quantify the degradation of the standard BAO reconstruction due to shot noise. The shot noise of power spectrum is calculated as $P_{\mathrm{shot}} = 1/\bar{n}$, where $\bar{n}$ is the galaxy number density in each redshift bin and calculated using the surface density given in Ref.~\cite{DESI:forecast}. The covariance matrix is then given by the inverse of the Fisher matrix, with forecast errors of measurements listed in Table~\ref{tab:desi-forecast-errors}.

\begin{table}[!htbp]
    \centering
    \begin{tabular}{ccccc}
        \toprule
        Tracer & $z_{\mathrm{eff}}$ & $\sigma(D_M/r_d)$ & $\sigma(D_H/r_d)$ & $\sigma(f\sigma_8)$ \\
        \midrule
        BGS & $0.20$ & $0.04790$ & $0.43217$ & $0.02273$ \\
        LRG & $0.75$ & $0.05756$ & $0.10548$ & $0.00969$ \\
        ELG & $1.35$ & $0.13397$ & $0.08649$ & $0.00612$ \\
        QSO & $1.85$ & $0.43943$ & $0.17484$ & $0.01872$ \\
        \bottomrule
    \end{tabular}
    \caption{Forecast errors of the BAO and RSD measurements in different tracers derived from the full DESI Fisher matrix.}
    \label{tab:desi-forecast-errors}
\end{table}

\subsubsection{CSST-like Weak Lensing Data}
\label{sec:WL data}

The weak lensing angular power spectrum synthetic data are also generated using the Boltzmann code \texttt{CAMB} and nonlinear corrections from \texttt{HMCode}~\cite{Mead:2015hmcode, Mead:2020vgs} with \texttt{mead2020} version in the Planck 2018 $\Lambda$CDM cosmology. The data points are generated in terms of the cosmic shear power spectrum $C_{\kappa\kappa}^{ij}$, the galaxy clustering angular power spectrum $C_{gg}^{ij}$, and the angular galaxy-galaxy lensing power spectrum $C_{g\kappa}^{ij}$, where $i,j=1,\ldots,4$ denote the four tomographic redshift bins. For $C_{\kappa\kappa}^{ij}$, we retain all redshift bin combinations with $i \leq j$. For the $C_{g\kappa}^{ij}$, we retain the bin pairs with $i \leq j$, where $i$ labels the lensing galaxy bin and $j$ labels the source galaxy bin. For $C_{gg}^{ij}$, we include the four auto-bin spectra $(i,i)$ and the three cross spectra between adjacent redshift bins $(i,i+1)$. The data vector therefore contains 10 cosmic shear spectra, 10 galaxy--galaxy lensing spectra, and 7 galaxy clustering spectra. We construct the synthetic data as CSST-like data, which has the expected characteristics and statistical precision from the China Space Station Telescope (CSST) survey, a Stage-IV survey aiming at exploring a larger volume of the universe and giving more precise parameter constraints.

To model the angular power spectra and make them CSST-like data, we follow the method of Ref.~\cite{Xiong:CSST}. We use the Limber approximation~\cite{Limber:1954} to compute the angular power spectra, with the $\ell$ bins and scale cuts the same as those used in Ref.~\cite{Xiong:CSST}. The scale cut $\ell_{\max}=[425,789,1097,1523]$ is only applied to the galaxy clustering and galaxy-galaxy lensing power spectrum to remove scales where nonlinear galaxy bias matters, while retaining nonlinear effects in the matter power spectrum to test the emulator.

\begin{figure}[!htbp]
    \centering
    \includegraphics[width=0.72\textwidth]{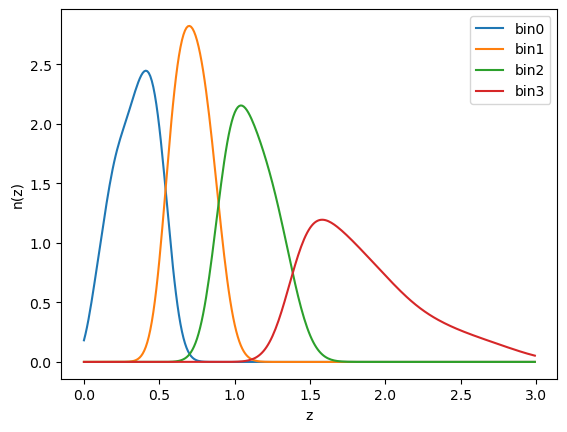}
    \caption{The redshift distribution $n(z)$ of the CSST survey used in the angular power spectrum calculation. Different colors represent different redshift bins. The $n(z)$ is generated from the JiuTian-1G simulation catalog and corresponds to the expected redshift distribution of the CSST survey.}
    \label{fig:nz}
\end{figure}

For the photometric redshift distribution $n(z)$ used in angular power spectrum calculation, we use the same $n(z)$ as in Ref.~\cite{Xiong:CSST} shown in Fig.~\ref{fig:nz}, which is generated from the JiuTian-1G simulation catalog and corresponds to the expected redshift distribution of the CSST survey. When modelling the redshift distribution $n(z)$ of each bin, we use the same method as shown in Ref.~\cite{Xiong:CSST} and DES-Y3 paper~\cite{DES:Y3Maglim, DES:Y3calibration}, introducing shift parameters $\Delta z^i$ and stretch parameters $\sigma_{\mathrm{z}}^i$ to model the uncertainty of the redshift distribution. Although the current version of \texttt{ReACT} is calibrated for $z\leq2.5$, and the effective redshift in the CSST-like redshift distribution extends to $z\simeq3$, the angular power spectra receive only a subdominant contribution from the high-redshift part given that most of the MG nonlinear corrections take place at redshifts below $z=2.5$, and nonlinear corrections become progressively smaller at high redshift as the density field approaches the linear regime. Motivated by this, we follow Ref.~\cite{Wang:2024mge} and use $P^{\mathrm{pseudo}}(k,z)$ at $z>2.5$ as an approximation, which has been implemented in \texttt{MGCAMB}. We therefore keep the full redshift distribution in this synthetic data validation and treat the emulator prediction over $2.5<z\leq3$ as a controlled approximation.

In galaxy clustering, the galaxy bias is modelled as a linear bias with one free parameter $b_i$ for each redshift bin in the angular power spectrum. We fix the $b_i = [1.075, 1.253, 1.535, 2.425]$ to generate the synthetic data. In cosmic shear, we introduce only the multiplicative shear-calibration parameters $m_i$, which are set to zero in the fiducial model. We also include intrinsic alignments using the NLA model widely adopted in DES MG analyses. This model has two free parameters, $A_{\mathrm{IA}}$ and $\alpha_{\mathrm{IA}}$, which are the amplitude and redshift evolution of the intrinsic alignment.

\begin{figure}[!htbp]
    \centering
    \includegraphics[width=1\textwidth]{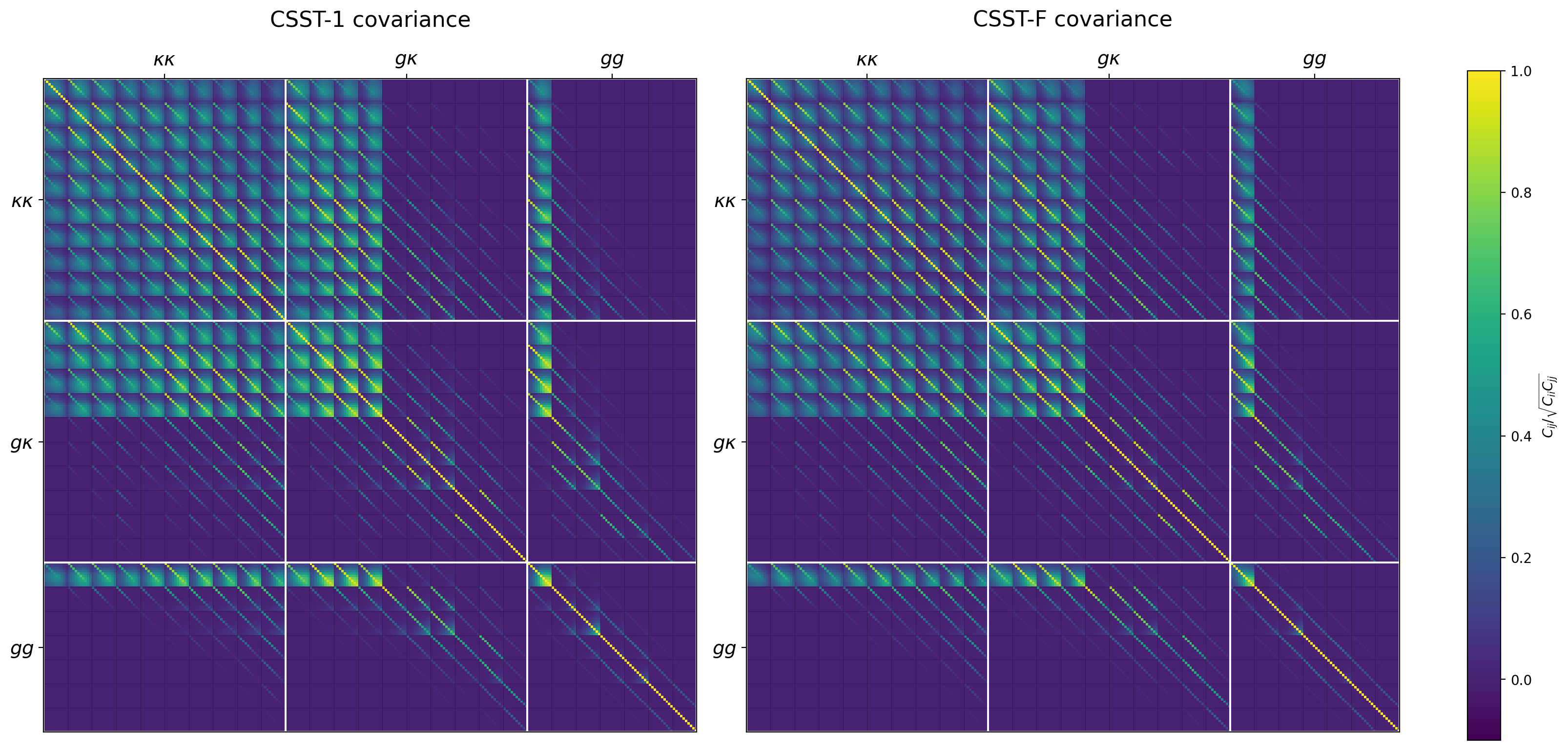}
    \caption{The covariance matrices of the CSST-like weak lensing angular power spectrum synthetic data. The matrices include both Gaussian and non-Gaussian contributions. The left one is calculated for a sky area of 1750 $\mathrm{deg}^2$, which is one tenth of the full CSST survey area, as a prediction of CSST first-year results. The right one is calculated for a sky area of 17500 $\mathrm{deg}^2$, which is the full CSST survey area, as a prediction of the CSST final results.}
    \label{fig:wl-cov}
\end{figure}

For the covariance matrix of the angular power spectrum synthetic data, we calculate the theoretical covariance matrix at the fiducial $\Lambda$CDM cosmology, consisting of Gaussian and non-Gaussian contributions. The Gaussian covariance is computed following Ref.~\cite{Xiong:CSST}, including the shot-noise terms from the mock. For the non-Gaussian covariance, following the same treatment as Ref.~\cite{Xiong:CSST}, we include two halo-model contributions: the super-sample covariance (SSC)~\cite{Takada:SSC} and the connected non-Gaussian covariance (cNG)~\cite{Takada:cNG}. These terms are computed with the implementation in \texttt{CCL}\footnote{\url{https://github.com/LSSTDESC/CCL}}~\cite{CCL}. The halo-model trispectrum used in the \texttt{CCL} covariance calculations is generated with the NFW halo profile~\cite{Navarro:1996NFW}, the Duffy concentration--mass relation~\cite{Duffy:2008}, the Tinker halo mass function~\cite{Tinker:2008}, and the Tinker halo-bias model~\cite{Tinker:2010}. Here we generate the covariance in two cases: for the one tenth CSST survey sky area, i.e. 1750 $\mathrm{deg}^2$, as a prediction of CSST first-year results (CSST-I), and for the full CSST survey sky area, i.e. 17500 $\mathrm{deg}^2$, as a prediction of CSST final results (CSST-F). The normalized covariance matrices of two cases are shown in Fig.~\ref{fig:wl-cov}.

\subsubsection{Parameter Prior}

In the MCMC analysis, we vary the following parameters:
\begin{itemize}
    \item Cosmological parameters: $\log A$ ($\ln(A_s\times10^{10})$), $n_s$, $H_0$, $\Omega_b h^2$, $\Omega_c h^2$, $\tau$
    \item MG parameters: $\mu_1,\ldots,\mu_{11}$, $\Sigma_1,\ldots,\Sigma_{11}$, $\Omega_{X,1},\ldots,\Omega_{X,10}$, $p_1$
    \item Nuisance parameters: mentioned in Section~\ref{sec:WL data}, including $\Delta z^i$, $\sigma_{\mathrm{z}}^i$, $b_i$, $m_i$, $A_{\mathrm{IA}}$ and $\alpha_{\mathrm{IA}}$ for weak lensing likelihood
\end{itemize}

The priors on the parameters used in the MCMC analysis are shown in Table~\ref{tab:prior}. The cosmological parameters are given Gaussian priors based on the Planck 2018 results~\cite{Planck:2018Planck}, considering an approximation of the external background and primordial amplitude information from CMB, while the MG parameters are given wide uniform priors to allow for a broad exploration of the parameter space. The nuisance parameters are given priors consistent with Ref.~\cite{Xiong:CSST}. The neutrino mass is fixed to $\sum m_\nu=0.06\,\mathrm{eV}$ in \texttt{MGCAMB}, and the intrinsic-alignment pivot redshift is fixed to $z_{0,\mathrm{IA}}=0.62$.

\begin{table}[t]
    \centering
    \begin{tabular}{ll}
        \toprule
        Parameter & Prior \\
        \midrule
        $\log A$ & $\mathcal{N}(3.036394,0.014)$ \\
        $n_s$ & $\mathcal{N}(0.9649,0.0042)$ \\
        $H_0$ & $\mathcal{N}(67.36,0.54)$ \\
        $\Omega_b h^2$ & $\mathcal{N}(0.02237,0.00015)$ \\
        $\Omega_c h^2$ & $\mathcal{N}(0.120,0.0012)$ \\
        $\tau$ & $\mathcal{N}(0.0544,0.0073)$ \\
        \midrule
        $\mu_i$, $i=1,\ldots,11$ & $\mathcal{U}(0,2)$ \\
        $\Sigma_i$, $i=1,\ldots,11$ & $\mathcal{U}(0,2)$ \\
        $\Omega_{X,1}$ & $\mathcal{U}(-2,4)$ \\
        $\Omega_{X,i}$, $i=2,\ldots,10$ & $\mathcal{U}(-1,3)$ \\
        $p_1$ & $\mathcal{U}(-2,2)$ \\
        \midrule
        $\Delta z^i$, $i=1,\ldots,4$ & $\mathcal{N}(0,0.01)$ \\
        $\sigma_{\mathrm{z}}^i$, $i=1,\ldots,4$ & $\mathcal{N}(1,0.05)$ \\
        $b_i$, $i=1,\ldots,4$ & $\mathcal{U}(0,5)$ \\
        $m_i$, $i=1,\ldots,4$ & $\mathcal{N}(0,0.01)$ \\
        $A_{\mathrm{IA}}$ & $\mathcal{U}(-5,5)$ \\
        $\alpha_{\mathrm{IA}}$ & $\mathcal{U}(-5,5)$ \\
        \bottomrule
    \end{tabular}
    \caption{Priors used in the $\Lambda$CDM synthetic-data MCMC validation. Here $\mathcal{U}(a,b)$ denotes a uniform prior over $[a,b]$, and $\mathcal{N}(\mu,\sigma)$ denotes a Gaussian prior with mean $\mu$ and standard deviation $\sigma$. The neutrino mass is fixed to $\sum m_\nu=0.06\,\mathrm{eV}$, $\Omega_{X,11}$ is set to the value of $\Omega_\Lambda$ in \texttt{MGCAMB}, and the intrinsic-alignment pivot redshift is fixed to $z_{0,\mathrm{IA}}=0.62$.}
    \label{tab:prior}
\end{table}

\subsection{Validation outside the training ranges}

The parameter priors used in the MCMC analyses extend beyond some of the training ranges listed in Table~\ref{tab:training-params}. For parameter values outside these ranges, predictions are extended and obtained by the trained \texttt{CosmoPower} network. Our fiducial cosmology is the Planck 2018 $\Lambda$CDM cosmology, which is within the training ranges. However, our training ranges refer to the 3$\sigma$ ranges of the chains from Ref.~\cite{Pogosian:2021mio}, in which the posterior distributions of three reconstructed functions are not centralized at the fiducial cosmology. Therefore, the MCMC chains can sample points outside the training ranges. We therefore examine the accuracy of these predictions outside the training regions sampled by the chains. We randomly select 2000 out-of-range parameter points from the chains and compare the emulated nonlinear matter power spectra with the reference \texttt{MGCAMB} calculations. As in Section \ref{sec:set-comparison}, we compute the spectral differences $\Delta_P(k,z)$ and present the corresponding statistical results to assess the performance of the emulator when extrapolating beyond its training range.

\section{Validation Results}
\label{sec:validation-results}

\subsection{Representative Spectra Results}
\label{sec:spectra-test}

In this section, we present the results of spectrum comparison for the selected representative cosmologies. Here we choose the Planck 2018 $\Lambda$CDM cosmology and an MG model whose linear power spectrum at $z=0$ is close to the $\Lambda$CDM result.

\begin{figure}[!htbp]
    \centering
    \includegraphics[width=1\textwidth]{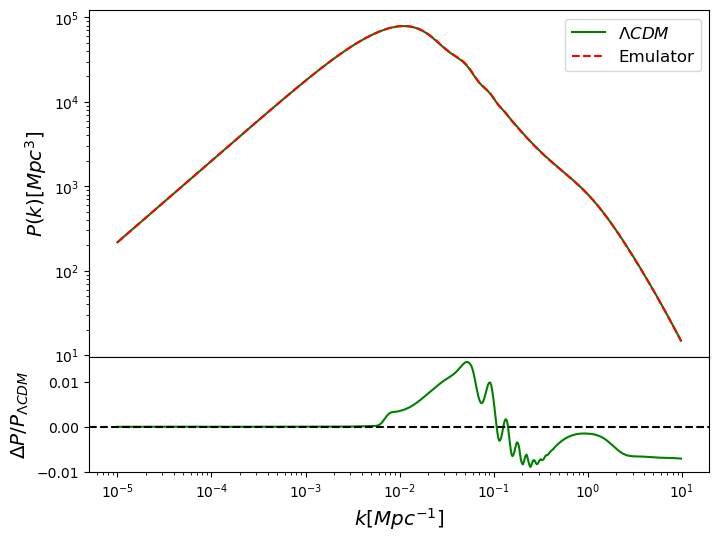}
    \caption{Validation of the emulator for the Planck 2018 $\Lambda$CDM cosmology at $z=0$. The upper panel compares the nonlinear matter power spectra predicted by the emulator (red dashed line) and the \texttt{MGCAMB} (green solid line). The lower panel shows the relative difference defined in Eq.~\eqref{eq:delta-p}. The emulator agrees with the reference result within $1.5\%$ over the scales shown.}
    \label{fig:validation-lcdm}
\end{figure}

The comparison for $\Lambda$CDM at $z=0$ is shown in Figure~\ref{fig:validation-lcdm}. The upper panel shows the direct comparison of the nonlinear power spectra between the emulator (red dashed line) and \texttt{MGCAMB} (green solid line), while the lower panel shows the relative difference of power spectra,
\begin{equation}
\Delta_P(k) = \frac{P_{\mathrm{emu}}^{\mathrm{NL}}(k,z=0)}{P_{\mathrm{MGCAMB}}^{\mathrm{NL}}(k,z=0)} - 1.
\label{eq:delta-p}
\end{equation}
The emulator prediction is consistent with the \texttt{MGCAMB} result within $1.5\%$ across all scales shown in the figure. The oscillation of the relative difference in the range $0.01 < k < 0.1$ $\mathrm{Mpc}^{-1}$ may be due to emulator error. In this range, the nonlinear correction starts to affect the power spectrum, and baryon acoustic oscillation (BAO) features also contribute. This makes it harder for the emulator to capture the power spectrum information and interpolate the power spectrum features. Therefore, the emulator shows an emulator error of about $1\%$, which is acceptable. The oscillation disappears at smaller scales, where the nonlinear correction becomes more significant and the power spectrum is smoother, making the emulator more accurate.

\begin{figure}[!htbp]
    \centering
    \includegraphics[width=0.9\textwidth]{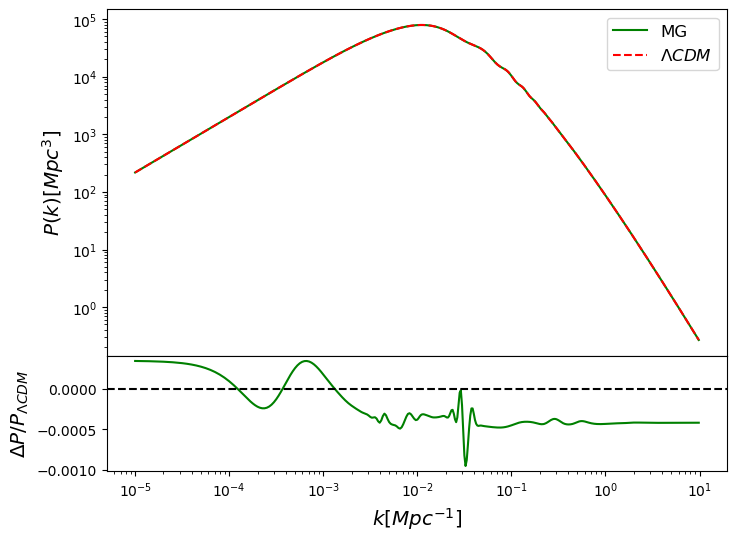}
    \caption{Linear matter power spectra at $z=0$ for the representative MG model (green solid line) and the $\Lambda$CDM model (red dashed line). The upper panel compares the two spectra, and the lower panel shows their relative difference. The representative MG model is selected such that its linear spectrum differs from the $\Lambda$CDM result by less than $0.1\%$ over the scales shown.}
    \label{fig:validation-mg-linear}
\end{figure}

Following the representative model selection described in Section~\ref{sec:spectra-comparison}, we vary only the $\mu$ nodes while fixing the other parameters to their $\Lambda$CDM values and GR limits. To get this representative MG model ($\mu$ set), we set a criterion that the linear power spectrum at $z=0$ should be within $0.1\%$ difference of the $\Lambda$CDM result. Finally, we find one $\mu$ set that satisfies this criterion:
    \begin{equation}
    \begin{aligned}
        \mu = [&1.0,\;1.018495483619,\;0.936640929860,\;1.144963586522,\;0.899900000000,\\
                 &1.103296381305,\;0.940339137892,\;0.960162176158,\;0.904265634155,\\
                 &1.153314438380,\;0.938622232110]
    \end{aligned}
    \end{equation}
and we use it for the validation test. The comparison of the linear power spectrum between this MG model and the $\Lambda$CDM model is shown in Figure~\ref{fig:validation-mg-linear}, indicating that these two models are in good agreement in linear theory prediction at $z=0$.

\begin{figure}[!htbp]
    \centering
    \includegraphics[width=0.9\textwidth]{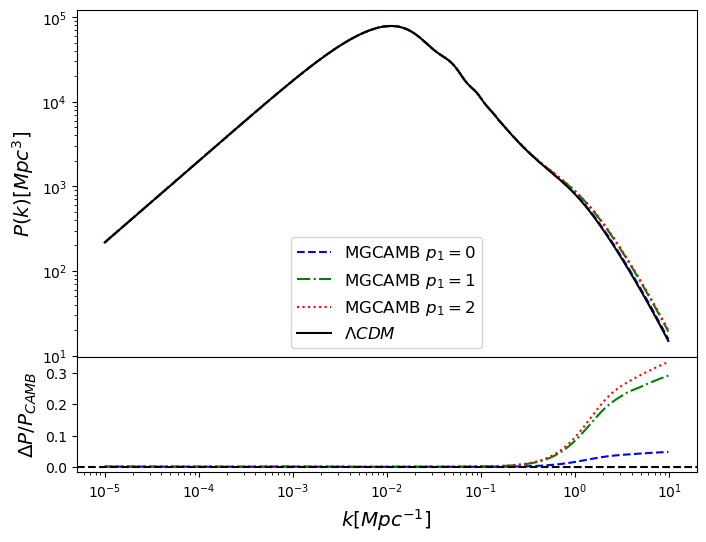}
    \caption{Nonlinear matter power spectra for the representative MG model with different values of $p_1$. The $\Lambda$CDM result is shown as the black solid line, while the MG results with $p_1=0$, $p_1=1$, and $p_1=2$ are shown as the blue dashed, green dash-dotted, and red dotted lines, respectively. The lower panel shows the relative difference with respect to $\Lambda$CDM. Larger values of $p_1$ lead to a stronger nonlinear MG correction at small scales.
    }
    \label{fig:validation-mg-nonlinear}
\end{figure}

To compare the nonlinear power spectrum of the selected MG model with the $\Lambda$CDM result, we also vary the $p_1$, which controls the strength of screening mechanism phenomenologically, to show the effect of MG nonlinear correction. A larger $p_1$ corresponds to a weaker screening mechanism and thus a stronger MG nonlinear effect at small scales. Therefore, by varying $p_1$, we can show how the MG nonlinear effect changes with different strengths of the screening mechanism, which is shown in Figure~\ref{fig:validation-mg-nonlinear}. As above, the upper panel is the nonlinear power spectra in different models and the lower panel is their relative difference. In the figure, the relative difference increases as the $k$ increases. Relative to the black solid $\Lambda$CDM curve, the MG model with $p_1=0$ (blue dashed line) shows a deviation up to $4\%$ at small scales, much smaller than the deviation of the MG model with $p_1=1$ (green dashdot line) which reaches around $30\%$, and the MG model with $p_1=2$ (red dotted line) shows an even larger deviation exceeding $30\%$. This indicates that the MG nonlinear effect becomes stronger at smaller scales and with larger $p_1$, and nonlinear power spectra can differ significantly between theories even though they can predict similar linear power spectra.

\begin{figure}[!htbp]
    \centering
    \makebox[\textwidth][c]{\includegraphics[width=1.08\textwidth]{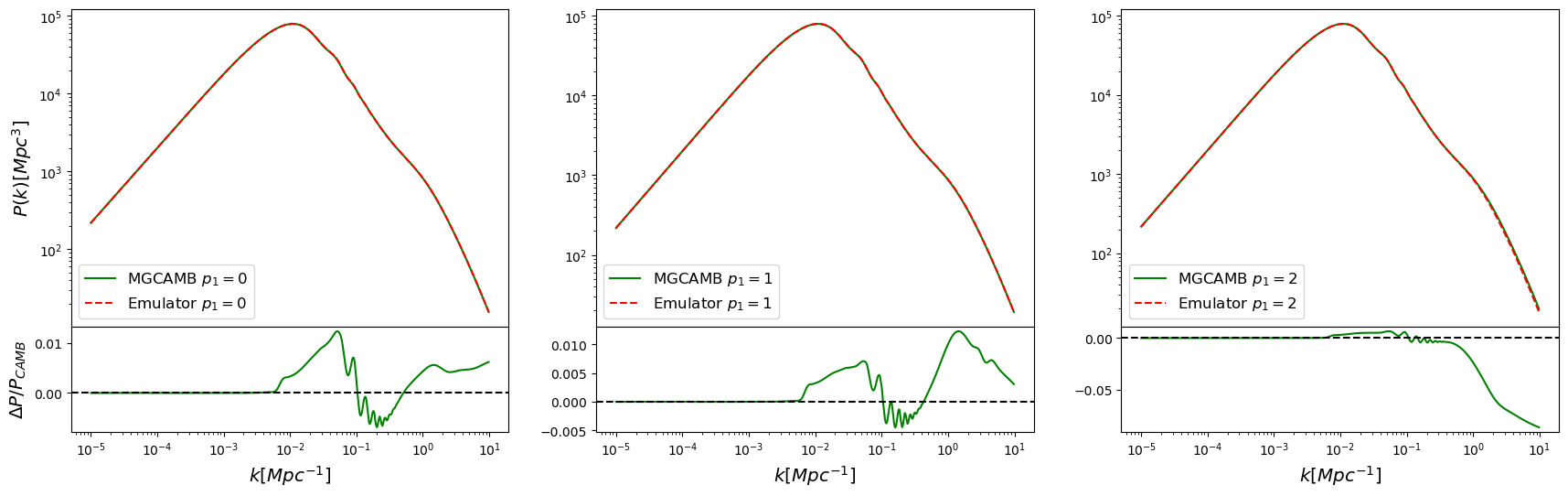}}
    \caption{Validation of the emulator for the representative MG models with $p_1=0$, $p_1=1$, and $p_1=2$ from left to right. In each panel, the upper subpanel compares the nonlinear matter power spectra from the emulator and the \texttt{MGCAMB}, while the lower subpanel shows the relative difference. The emulator agrees with the reference result within $1.5\%$ over the scales shown for $p_1=0$ and $p_1=1$, and within $1.5\%$ for $k<1\,\mathrm{Mpc}^{-1}$ for $p_1=2$.}
    \label{fig:validation-mg}
\end{figure}

The emulator validation result for the MG models is shown in Figure~\ref{fig:validation-mg}. As before, the upper panel shows the comparison of the nonlinear power spectra and the lower panel shows the relative difference. From the left panel to the right, we show the results for the MG model with $p_1=0,1,2$. In the figure, for $p_1=0$ and $p_1=1$ cases, the emulator predictions are consistent with the \texttt{MGCAMB} results within $1.5\%$ across all scales, as the emulator error in the range $0.01 < k < 0.1$ still exists as in the $\Lambda$CDM case. For the $p_1=2$ case, the emulator prediction is still consistent with the \texttt{MGCAMB} results within $1.5\%$ at $k<1$ $\mathrm{Mpc}^{-1}$, but it shows a larger deviation at smaller scales and reached around $7\%$ at $k=5$ $\mathrm{Mpc}^{-1}$, which is possibly due to the fact that with a higher $p_1$ value, the weaker screening mechanism results in the more significant MG nonlinear effect at small scales, making it harder for the emulator to capture this effect accurately.

\subsection{Validation Set Results}
\label{sec:validation-set}

\begin{figure}[!htbp]
    \centering
    \includegraphics[width=0.9\textwidth]{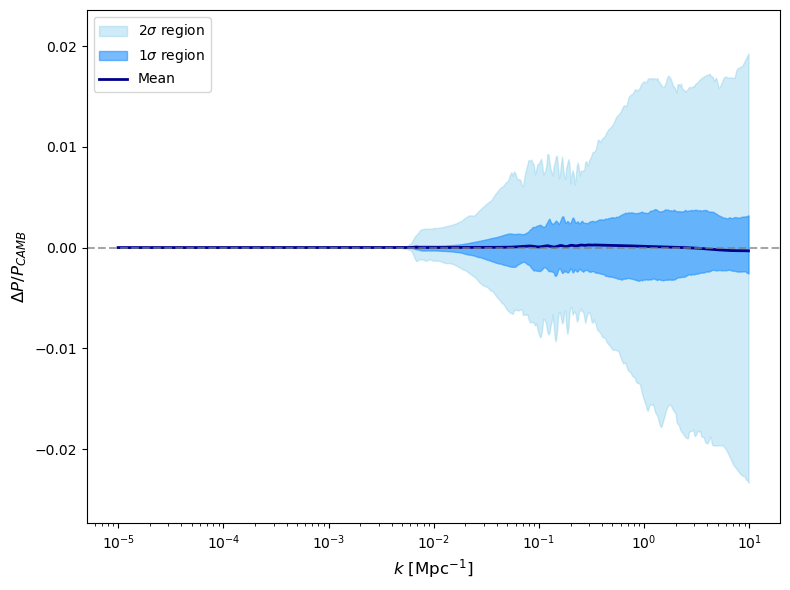}
    \caption{Mean residual and $1\sigma$ and $2\sigma$ residual bands of $\Delta_P(k,z)$ over the 2000 independent validation points. The dark blue line shows the mean results, the deep and light blue regions show the $1\sigma$ and $2\sigma$ bands, respectively, and the dashed black line denotes $\Delta_P=0$. The $2\sigma$ residual remains below $1\%$ for $k<0.5\,\mathrm{Mpc}^{-1}$ and reaches approximately $2\%$ at smaller scales.}
    \label{fig:validation-band}
\end{figure}

Using the 2000 validation points described in Section~\ref{sec:set-comparison}, we compute the difference in Eq.~\eqref{eq:delta-p} between the emulator prediction and the \texttt{MGCAMB} result for each point. Then we calculate the mean, $1\sigma$ and $2\sigma$ bands of $\Delta_P(k,z)$ of the points in the validation set at each $k$. The mean value, $1\sigma$ and $2\sigma$ bands of $\Delta_P(k,z)$ are shown in Figure~\ref{fig:validation-band}, in which the dark blue line is the mean of deviations, the deep blue band shows the $1\sigma$ uncertainty, the light blue band shows the $2\sigma$ uncertainty, and the dashed line represents the fiducial value $\Delta_P(k,z)=0$. In the figure, the deviation starts to appear at $k\sim0.01$ $\mathrm{Mpc}^{-1}$, which is the scale where the nonlinear correction starts to affect the power spectrum, showing the emulation error of the emulator prediction. We do not expect significant emulation errors at $k\leq0.01$ $\mathrm{Mpc}^{-1}$ as spectra here are dominated by linear perturbation theory and we do not expect any significant deviations. The mean value of $\Delta_P(k,z)$ is close to zero across all scales, while the $2\sigma$ band of $\Delta_P(k,z)$ is within $1\%$ at $k<0.5$ $\mathrm{Mpc}^{-1}$ and increases to about $2\%$ at smaller scales, which shows the statistical accuracy of the emulator in our parameter space. The fact that the relative difference increases with decreasing scale is due to the strong and increasing nonlinear effects at small scales. Similar trends can be found in the result shown in the right panel of Figure~\ref{fig:validation-mg}, indicating that while the MG model brings significant nonlinear effects which become stronger with larger $k$, the emulator finds it harder to capture the effects accurately, resulting in a larger emulating error. Nevertheless, the figure indicates that the emulator remains sufficiently accurate across all scales. As we demonstrate below, this level of accuracy is adequate for the tests considered in this work and for parameter inference with future Stage IV surveys.

\subsection{MCMC Validation Results}
\label{sec:mcmc-validation}

\begin{figure}[!htbp]
    \centering
    \includegraphics[width=1\textwidth]{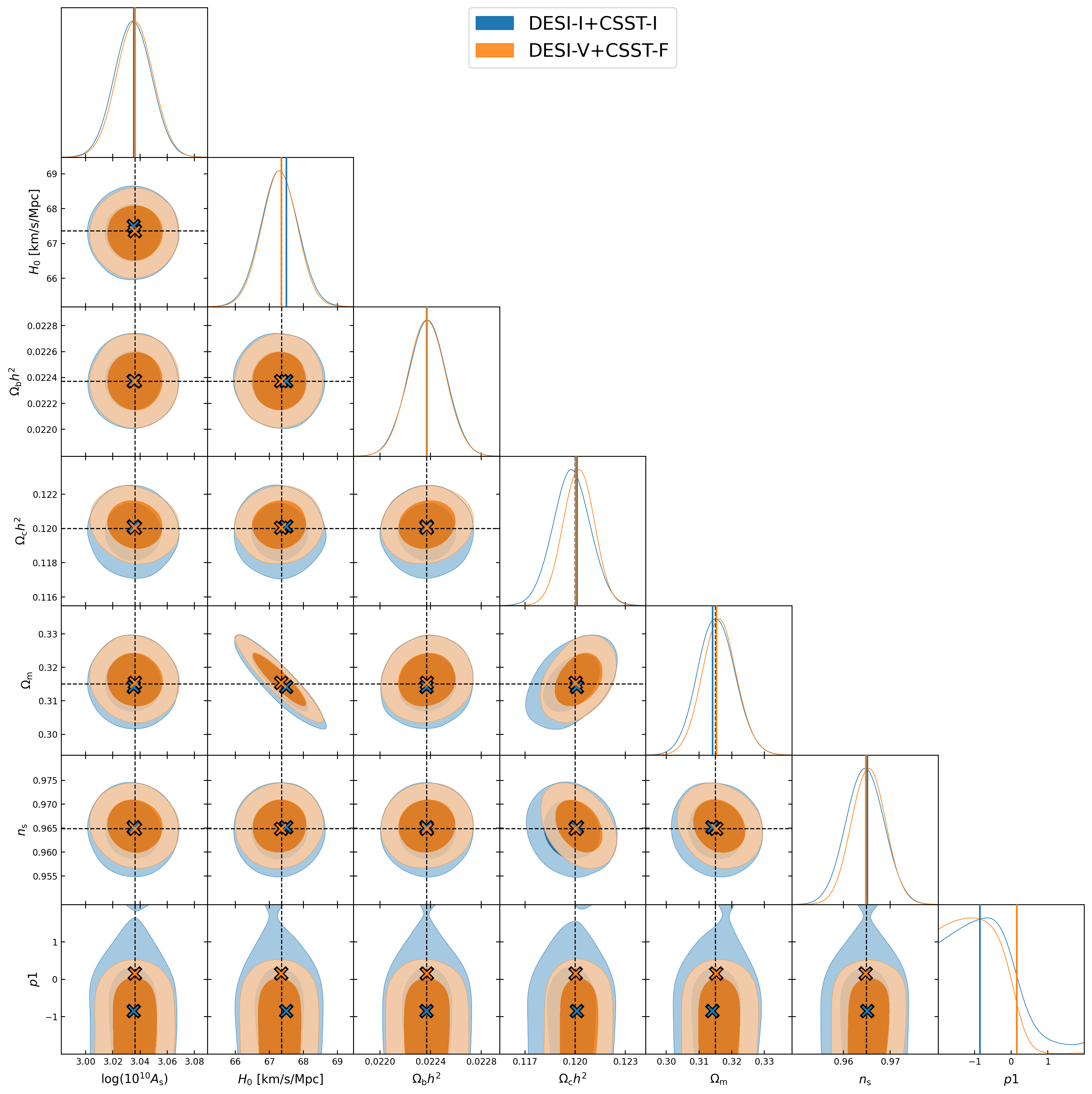}
    \caption{MCMC validation results for the cosmological parameters using $\Lambda$CDM synthetic data and CMB-informed priors. Blue contours show the DESI-I+CSST-I case, and orange contours show the DESI-V+CSST-F case. The contours indicate the $1\sigma$ and $2\sigma$ regions, the coloured crosses mark the best-fit points, and the black dashed lines show the fiducial values. The fiducial cosmology is recovered within the posterior uncertainties in both synthetic-data combinations.}
    \label{fig:cosmo_compare}
\end{figure}

Following the synthetic-data setup described in Section~\ref{sec:mcmc}, Figure~\ref{fig:cosmo_compare} shows the MCMC validation results of cosmological parameters using the $\Lambda$CDM synthetic data and CMB-informed priors. The contours are cosmological parameter results with their $1\sigma$ and $2\sigma$ regions. The blue contours represent DESI-I+CSST-I results and the orange contours represent the DESI-V+CSST-F results. The blue and orange solid lines are parameter best-fit values in different data, with the same color cross in contour, while the black dashed lines represent their fiducial values. The figure shows that the peaks of the parameters posterior distribution and the crosses are both close to the fiducial values. We do not find large differences between the parameter errors of the two data combinations, indicating that even if we have more precise Stage-IV galaxy survey results, CMB-informed prior (or CMB observation result) still dominates the errors of these parameters. These results indicate that with the emulator predictions in the likelihood, we can recover the fiducial values of cosmological parameters very well in both mean and best-fit values within the uncertainties of the synthetic data.

\begin{figure}[!htbp]
    \centering
    \includegraphics[width=1\textwidth]{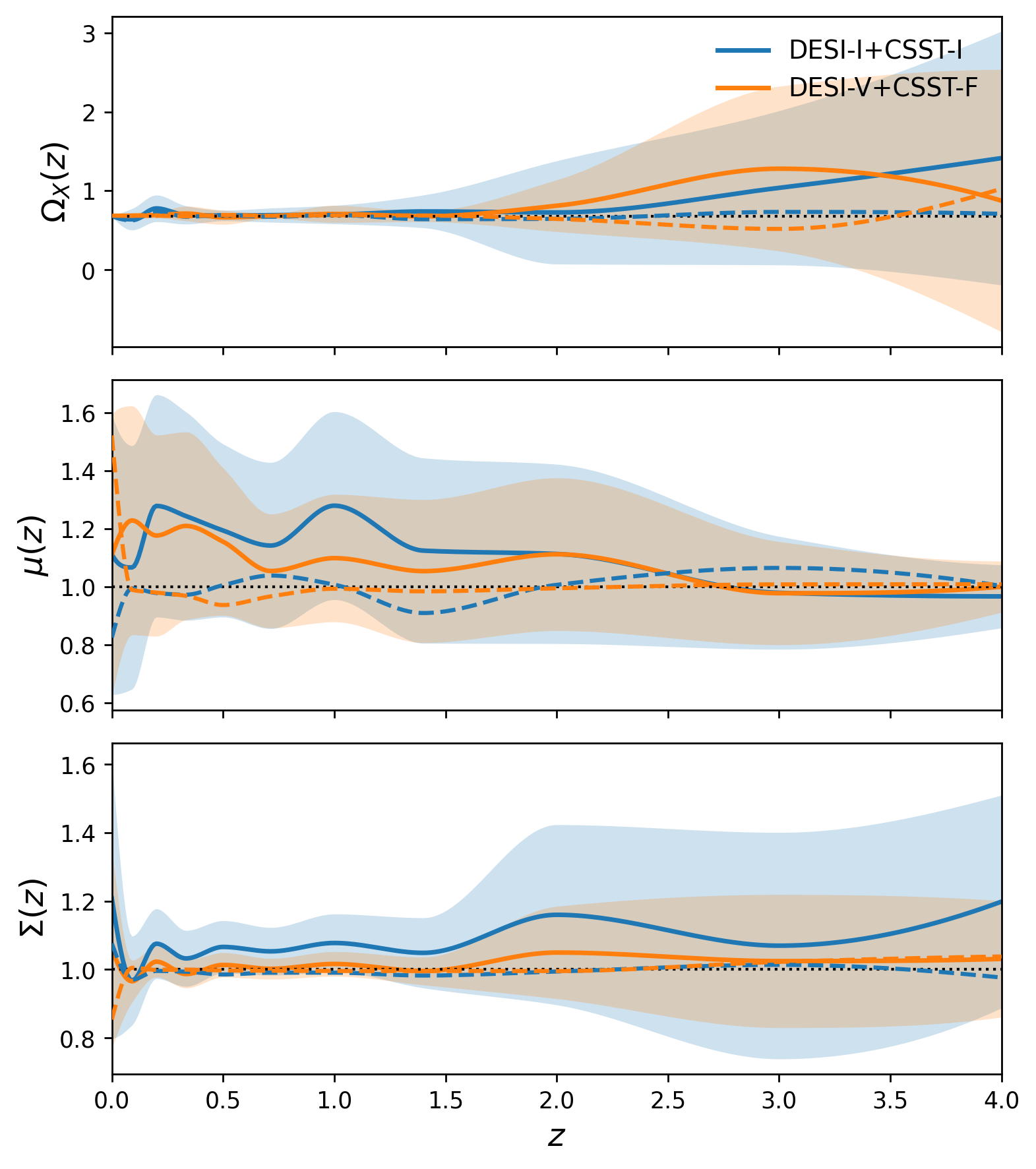}
    \caption{MCMC validation results for the three reconstructed MG functions. Blue and orange results correspond to DESI-I+CSST-I and DESI-V+CSST-F, respectively. Solid lines show posterior means, dashed lines show best-fit values, and shaded regions show the $1\sigma$ uncertainties. The black dotted lines denote the fiducial values, corresponding to the GR limits for $\mu$ and $\Sigma$ and the $\Lambda$CDM fiducial background evolution for $\Omega_X$. The reconstructed functions are consistent with their fiducial values within the statistical uncertainties.}
    \label{fig:mg_compare}
\end{figure}

Figure~\ref{fig:mg_compare} shows the MCMC validation results for the three reconstructed MG functions sampled at 11 redshift nodes. The blue results correspond to DESI-I+CSST-I, while the orange results correspond to DESI-V+CSST-F. The solid lines show the posterior mean, the dashed lines show the best-fit values, and the shaded regions indicate the $1\sigma$ uncertainty. The black dotted lines denote the $\Lambda$CDM fiducial values, corresponding to the GR limits for $\mu$ and $\Sigma$ and the fiducial background evolution for $\Omega_X$. The corresponding numerical results are summarized in Table~\ref{tab:mcmc-bestfit}, where each entry gives the posterior mean followed by the best-fit value in parentheses. The total likelihood chi-square is defined as the sum of the BAO+RSD and CSST-like $3\times2$pt contributions for the best-fit result and does not include the prior contribution.

For both data combinations, the best-fit values of $\mu$ and $\Sigma$ are close to their GR limits, except at very low redshift. This low-redshift deviation is expected because the corresponding nodes have limited support from the angular clustering and lensing power spectrum, and are mainly constrained by the first tomographic bin. Additionally, high redshift data cannot give any constraints on MG parameters at low redshift. As a result, these nodes are weakly constrained and are more sensitive to projection effects and parameter degeneracies. For DESI-V+CSST-F, the posterior means of $\Sigma$ are still consistent with GR. However, the posterior means of $\mu$ for DESI-V+CSST-F, and of both functions for DESI-I+CSST-I, show deviations from the GR limits, but remain within $1\sigma$ uncertainty.

\begin{figure}[!htbp]
    \centering
    \includegraphics[width=1\textwidth]{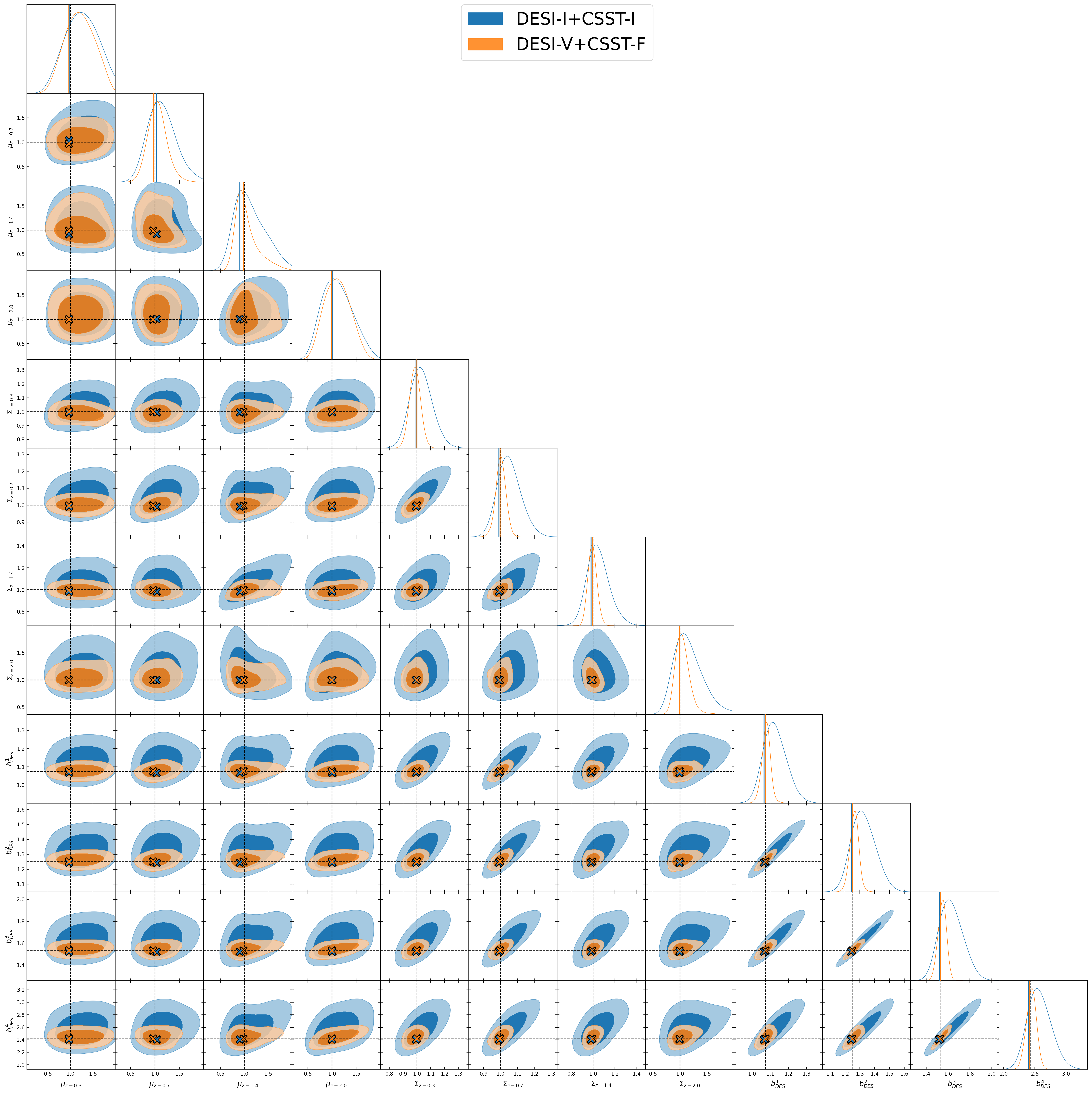}
    \caption{Degeneracies between selected MG parameters and linear galaxy bias parameters. The colour follows Figure~\ref{fig:cosmo_compare}. The degeneracy between $\Sigma_i$ and $b_i$ is driven by $C_\ell^{\mathrm{g}\kappa}\propto b_i\Sigma_i P_{\delta\delta}$, while $\mu_i$ is correlated with $b_i$ through its effect on the growth and amplitude of matter clustering. The degeneracies are reduced in the higher-precision DESI-V+CSST-F case.}
    \label{fig:projection-effect}
\end{figure}

To explain the origin of these deviations, Figure~\ref{fig:projection-effect} shows the degeneracy between selected MG nodes and the linear galaxy bias parameters $b_i$. We choose four MG nodes whose redshifts are close to the centres of the photometric redshift bins. In linear perturbation theory, the angular galaxy-galaxy lensing power spectrum can be approximately expressed as
\begin{equation}
    C_\ell^{\mathrm{g}\kappa} \propto b_i \Sigma_i P_{\delta\delta},
\end{equation}
which explains the strong degeneracy between $\Sigma_i$ and $b_i$. The $\mu$ parameters affect the growth of matter perturbations and also change the amplitude of the galaxy clustering, producing a weaker but still significant degeneracy with $b_i$. These degeneracies can shift the posterior means of the reconstructed MG functions away from their fiducial values without indicating a failure of the emulator, causing the projection effects observed in the MCMC results.

With the improved precision of DESI-V+CSST-F, the reconstructed MG functions and galaxy bias parameters move closer to their fiducial values. Together with the degeneracies shown in Figure~\ref{fig:projection-effect}, this behaviour is consistent with the projection effects contributing to the posterior mean shifts. The reconstructed $\Omega_X$ function nodes are also consistent with the fiducial $\Lambda$CDM background in both data combinations.

\begin{table}[!htbp]
    \centering
    \small
    \begin{tabular}{lccc}
        \toprule
        Parameter & $\Lambda$CDM fiducial & DESI-I+CSST-I & DESI-V+CSST-F \\
        \midrule
        $\log A$          & $3.036394$ & $3.034965\,(3.035459)$ & $3.036230\,(3.036317)$ \\
        $n_s$             & $0.964900$ & $0.964664\,(0.965130)$ & $0.965438\,(0.964816)$ \\
        $H_0$             & $67.360$   & $67.299\,(67.499)$     & $67.298\,(67.352)$ \\
        $\Omega_b h^2$    & $0.0223700$ & $0.0223744\,(0.0223696)$ & $0.0223718\,(0.0223704)$ \\
        $\Omega_c h^2$    & $0.120000$ & $0.119798\,(0.120106)$ & $0.120219\,(0.120029)$ \\
        $\tau$             & $0.054400$ & $0.054124\,(0.053630)$ & $0.054312\,(0.054483)$ \\
        $\chi^2_{\mathrm{tot}}$ & --- & $0.24824$ & $0.36563$ \\
        \bottomrule
    \end{tabular}

    \vspace{0.7em}
    \scriptsize
    \setlength{\tabcolsep}{2.2pt}
    \begin{tabular*}{\textwidth}{@{\extracolsep{\fill}}lrrr lrrr lrrr@{}}
        \toprule
        \multicolumn{4}{c}{$\mu(a)$ reconstruction} &
        \multicolumn{4}{c}{$\Sigma(a)$ reconstruction} &
        \multicolumn{4}{c}{$\Omega_X(a)$ reconstruction and $p_1$} \\
        \cmidrule(lr){1-4}\cmidrule(lr){5-8}\cmidrule(lr){9-12}
        Parameter & Fid. & I+I & V+F &
        Parameter & Fid. & I+I & V+F &
        Parameter & Fid. & I+I & V+F \\
        \midrule
        $\mu_1$    & $1$ & \shortstack{$0.96685$\\($1.00266$)} & \shortstack{$1.00029$\\($1.00870$)} &
        $\Sigma_1$ & $1$ & \shortstack{$1.19864$\\($0.97656$)} & \shortstack{$1.03107$\\($1.03828$)} &
        $\Omega_{X,1}$ & $0.6847$ & \shortstack{$1.41415$\\($0.70856$)} & \shortstack{$0.87477$\\($1.02594$)} \\
        $\mu_2$    & $1$ & \shortstack{$0.97913$\\($1.06497$)} & \shortstack{$0.97775$\\($1.00825$)} &
        $\Sigma_2$ & $1$ & \shortstack{$1.06980$\\($1.01380$)} & \shortstack{$1.02475$\\($1.02243$)} &
        $\Omega_{X,2}$ & $0.6847$ & \shortstack{$1.03754$\\($0.73255$)} & \shortstack{$1.27999$\\($0.51769$)} \\
        $\mu_3$    & $1$ & \shortstack{$1.11311$\\($1.00665$)} & \shortstack{$1.11199$\\($0.99472$)} &
        $\Sigma_3$ & $1$ & \shortstack{$1.16006$\\($0.99418$)} & \shortstack{$1.04975$\\($0.99552$)} &
        $\Omega_{X,3}$ & $0.6847$ & \shortstack{$0.72615$\\($0.64489$)} & \shortstack{$0.81068$\\($0.64411$)} \\
        $\mu_4$    & $1$ & \shortstack{$1.12458$\\($0.90950$)} & \shortstack{$1.05387$\\($0.98425$)} &
        $\Sigma_4$ & $1$ & \shortstack{$1.04881$\\($0.98249$)} & \shortstack{$0.99567$\\($0.99610$)} &
        $\Omega_{X,4}$ & $0.6847$ & \shortstack{$0.73896$\\($0.64026$)} & \shortstack{$0.67706$\\($0.68784$)} \\
        $\mu_5$    & $1$ & \shortstack{$1.27940$\\($1.00733$)} & \shortstack{$1.09874$\\($0.99320$)} &
        $\Sigma_5$ & $1$ & \shortstack{$1.07769$\\($0.99086$)} & \shortstack{$1.01636$\\($0.99575$)} &
        $\Omega_{X,5}$ & $0.6847$ & \shortstack{$0.69867$\\($0.69417$)} & \shortstack{$0.70717$\\($0.69028$)} \\
        $\mu_6$    & $1$ & \shortstack{$1.14198$\\($1.03863$)} & \shortstack{$1.05436$\\($0.96695$)} &
        $\Sigma_6$ & $1$ & \shortstack{$1.05318$\\($0.99094$)} & \shortstack{$1.00214$\\($0.99666$)} &
        $\Omega_{X,6}$ & $0.6847$ & \shortstack{$0.69160$\\($0.67155$)} & \shortstack{$0.68179$\\($0.68429$)} \\
        $\mu_7$    & $1$ & \shortstack{$1.19392$\\($1.00549$)} & \shortstack{$1.15537$\\($0.93702$)} &
        $\Sigma_7$ & $1$ & \shortstack{$1.06616$\\($0.98551$)} & \shortstack{$1.01344$\\($0.99748$)} &
        $\Omega_{X,7}$ & $0.6847$ & \shortstack{$0.68674$\\($0.69172$)} & \shortstack{$0.66418$\\($0.69401$)} \\
        $\mu_8$    & $1$ & \shortstack{$1.24351$\\($0.97270$)} & \shortstack{$1.20974$\\($0.96878$)} &
        $\Sigma_8$ & $1$ & \shortstack{$1.03245$\\($0.99379$)} & \shortstack{$0.98687$\\($0.99979$)} &
        $\Omega_{X,8}$ & $0.6847$ & \shortstack{$0.69436$\\($0.67160$)} & \shortstack{$0.71201$\\($0.67744$)} \\
        $\mu_9$    & $1$ & \shortstack{$1.27865$\\($0.97832$)} & \shortstack{$1.17641$\\($0.98013$)} &
        $\Sigma_9$ & $1$ & \shortstack{$1.07567$\\($0.99595$)} & \shortstack{$1.02316$\\($0.99910$)} &
        $\Omega_{X,9}$ & $0.6847$ & \shortstack{$0.77727$\\($0.74189$)} & \shortstack{$0.69374$\\($0.68544$)} \\
        $\mu_{10}$ & $1$ & \shortstack{$1.06647$\\($0.99438$)} & \shortstack{$1.22873$\\($0.98917$)} &
        $\Sigma_{10}$ & $1$ & \shortstack{$0.96755$\\($0.97043$)} & \shortstack{$0.96563$\\($1.00520$)} &
        $\Omega_{X,10}$ & $0.6847$ & \shortstack{$0.63890$\\($0.62600$)} & \shortstack{$0.68489$\\($0.68389$)} \\
        $\mu_{11}$ & $1$ & \shortstack{$1.11210$\\($0.82502$)} & \shortstack{$1.11028$\\($1.52144$)} &
        $\Sigma_{11}$ & $1$ & \shortstack{$1.20830$\\($1.07649$)} & \shortstack{$1.06063$\\($0.85394$)} &
        $p_1$ & --- & \shortstack{$-0.70346$\\($-0.85364$)} & \shortstack{$-0.92701$\\($0.14969$)} \\
        \bottomrule
    \end{tabular*}
    \caption{Posterior means and best-fit values of the cosmological and MG reconstruction parameters in the two synthetic-data combinations. Each result is reported as the posterior mean, with the best-fit value in parentheses. In the lower part of the table, I+I and V+F denote DESI-I+CSST-I and DESI-V+CSST-F, respectively. The $\Lambda$CDM fiducial model has $\mu_i=\Sigma_i=1$ and $\Omega_{X,i}=\Omega_{\Lambda}=0.6847$ at every reconstruction node. The parameter $p_1$ has no unique $\Lambda$CDM fiducial value because it is inactive in GR. The reported $\chi^2_{\mathrm{tot}}$ is the total likelihood chi-square for the best-fit results.}
    \label{tab:mcmc-bestfit}
\end{table}

\subsection{Accuracy outside the training ranges}

\begin{figure}[!htbp]
    \centering
    \includegraphics[width=1\textwidth]{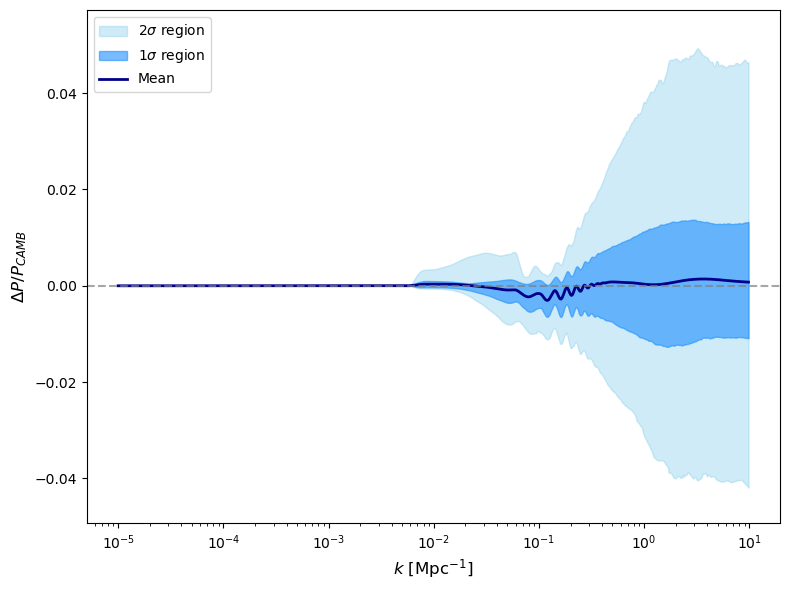}
    \caption{Mean residual and $1\sigma$ and $2\sigma$ regions of the nonlinear matter power spectrum predictions for 2000 randomly selected MCMC parameter points outside the training ranges. The dark blue line shows the mean, and the deep and light blue shaded regions show the $1\sigma$ and $2\sigma$ regions. The dashed line denotes zero residual.}
    \label{fig:outrange}
\end{figure}

Figure~\ref{fig:outrange} shows the residual distribution for the 2000 randomly selected out-of-range points. The mean residual (dark blue line) remains close to zero across the evaluated $k$ range, while the $1\sigma$ (deep blue) and $2\sigma$ (light blue) regions broaden towards smaller scales. Though the mean at BAO oscillation scales $k \simeq 0.1 $ $\mathrm{Mpc}^{-1}$ deviates from zero, the deviations are still within $1\sigma$ uncertainties. In the worst case, the $1\sigma$ region lies approximately within $1.5\%$, and the $2\sigma$ region increases roughly to $4\%$. Compared with the independent validation set in Section~\ref{sec:validation-set}, this sample exhibits a broader residual distribution at high $k$. These results characterise the extrapolation performance of the emulator in parameter space, demonstrating that it retains good accuracy even outside the region covered by the training set, although the errors are larger than those obtained within the training range.

Overall, all the validation tests show consistent performance of the emulator at different levels. The representative spectra tests demonstrate that the emulator reproduces the reference nonlinear power spectra accurately for $\Lambda$CDM and for selected MG models, with larger deviations appearing only in the stronger nonlinear MG case at high $k$. The validation-set test confirms that the deviations are statistically centred around zero over the parameter space, with the $2\sigma$ error remaining acceptable on the scales most relevant for weak lensing observations considered. The synthetic-data MCMC test shows that using the emulator inside the theory prediction pipeline recovers the input $\Lambda$CDM cosmology and the GR limits of the reconstructed MG functions within the posterior uncertainties. Finally, the out-of-range test shows that the emulator retains good accuracy beyond the parameter range covered by the training set, albeit with a broader distribution of residuals at high $k$.

\section{Application of the Emulator}
\label{sec:application}

The validated emulator can be used not only to accelerate the MG nonlinear power spectrum calculation, but also as a diagnostic tool for identifying MG nonlinear contributions in models with similar linear evolution. In this section we show three direct applications: identifying the MG-specific nonlinear signal, accelerating MCMC analyses with observational data, and forecasting the constraints from future galaxy surveys.

\subsection{Identifying MG Nonlinear Effects}
\label{sec:application-nonlinear-mg}

One useful application of the emulator is to predict the MG nonlinear effect from the input parameters without recomputing the full nonlinear power spectrum. As described in Section~\ref{sec:spectra-test}, different MG models can have similar linear matter power spectra at specific redshifts, leading to degeneracies at the level of linear perturbations. Although the growth of linear perturbations is similar, different structure formation histories behind them can lead to different nonlinear effects. Therefore, including nonlinear information can help break the degeneracy and identify the MG signal, which strengthens the test of deviations from GR. Since the emulator predicts the nonlinear correction $R_{\mathrm{MG}}$, given a set of input parameters whose linear predictions are close to each other, the emulator allows us to compare their nonlinear corrections directly and distinguish them through their different nonlinear effects. The purpose of this subsection is not to provide an additional accuracy validation of the emulator, which has been discussed in Section~\ref{sec:validation-results}. Instead, we use the emulator as a fast diagnostic tool to illustrate how nonlinear information can help distinguish MG models whose linear matter power spectra are nearly degenerate. In such cases, the nonlinear correction $R_{\mathrm{MG}}$ or the resulting nonlinear matter power spectrum can retain differences caused by the different growth histories, and therefore carries information that is not captured by the linear prediction alone.

\begin{figure}[htbp]
    \centering
    \includegraphics[width=1\textwidth]{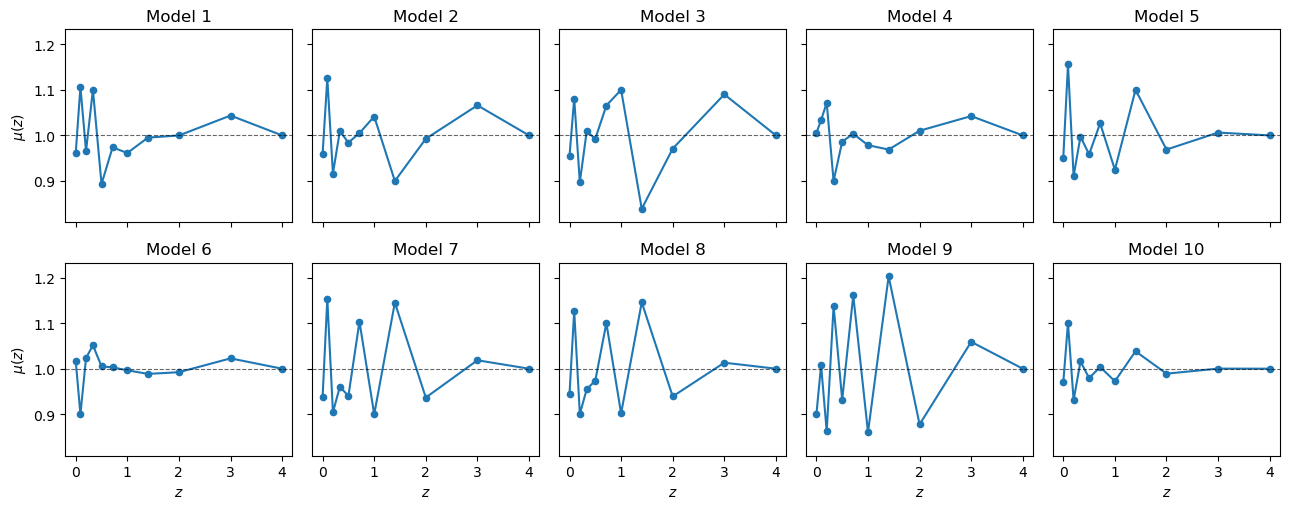}
    \caption{The ten MG models selected for the nonlinear effect prediction, whose linear power spectra are all within $0.1\%$ difference of the $\Lambda$CDM result at $z=0$.} 
    \label{fig:spectra-set}
\end{figure}

\begin{figure}[htbp]
    \centering
    \includegraphics[width=0.9\textwidth]{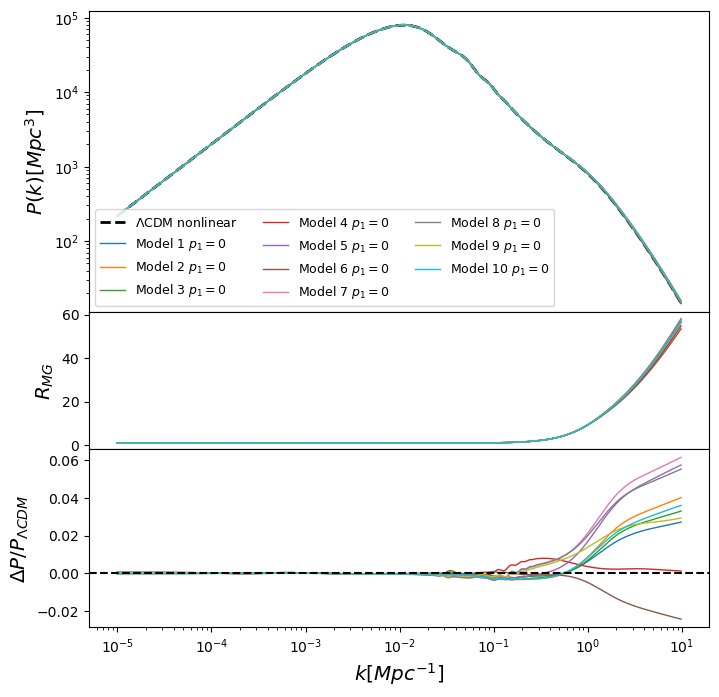}
    \caption{The MG nonlinear effect predicted by the emulator for the ten MG models selected as described in Section~\ref{sec:spectra-comparison} at $z=0$ with $p_1=0$. The upper panel shows the nonlinear power spectra in different selected models and $\Lambda$CDM, the middle panel shows the $R_{\mathrm{MG}}$, and the lower panel shows the relative difference of power spectra between MG models and $\Lambda$CDM. Both MG models and $\Lambda$CDM nonlinear power spectra are predicted by the emulator with \texttt{MGCAMB} linear power spectra.}
    \label{fig:mg-nonlinear-effect-p1-0}
\end{figure}

\begin{figure}[htbp]
    \centering
    \includegraphics[width=0.9\textwidth]{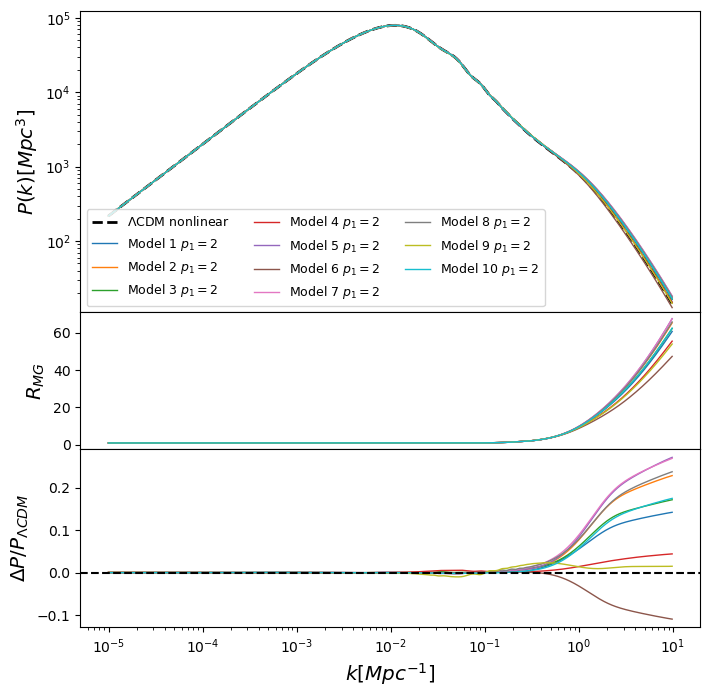}
    \caption{Same as Figure~\ref{fig:mg-nonlinear-effect-p1-0} but with $p_1=2$, showing a stronger MG nonlinear effect. Although the emulation error is larger than in the $p_1=0$ case, the emulator still captures the enhanced nonlinear response.}
    \label{fig:mg-nonlinear-effect-p1-2}
\end{figure}

As an example, we show the emulator results for the MG nonlinear effect using a total of ten MG models selected following the criteria shown in Section~\ref{sec:spectra-comparison}. Figure~\ref{fig:spectra-set} shows these ten selected MG models whose linear power spectra are all within $0.1\%$ difference of the $\Lambda$CDM result. The nonlinear power spectra of these models are shown at $z=0$ for $p_1=0$ in Figure~\ref{fig:mg-nonlinear-effect-p1-0} and for $p_1=2$ in Figure~\ref{fig:mg-nonlinear-effect-p1-2}, with different values of $p_1$ quantifying the strength of MG nonlinear effect. In both figures, the upper panel shows the nonlinear power spectra in different selected models and $\Lambda$CDM, the middle panel shows the $R_{\mathrm{MG}}$, and the lower panel shows the relative difference of power spectra between MG models and $\Lambda$CDM, which can also be considered as the relative difference of $R_{\mathrm{MG}}$ since the linear power spectra of these models are almost the same. All nonlinear power spectra including $\Lambda$CDM are predicted by the emulator with \texttt{MGCAMB} linear power spectra.

Considering the relative differences shown in Figure~\ref{fig:validation-mg}, although the deviation increases as the screening mechanism becomes weaker, GR is tightly tested in the Solar System and on small scales where nonlinear effects dominate, which means viable MG models are generally expected to recover GR efficiently in high-density or nonlinear regimes through screening mechanisms. As is shown in Section~\ref{sec:spectra-test}, the emulator result is consistent with the \texttt{MGCAMB} nonlinear extension result with a small $p_1$, which indicates that the emulator can still work effectively to help break the degeneracy of different models at the linear-perturbation level and identify the MG signal. We note that the emulation errors on BAO scales in $\Delta P/P_{\Lambda\mathrm{CDM}}$ shown in Figure~\ref{fig:validation-mg} are significantly reduced when the emulator prediction for $\Lambda$CDM is used in the denominator, as the emulation errors in the MG and $\Lambda$CDM power spectra cancel in the ratio. Additionally, to calculate the nonlinear power spectrum for these MG models, it costs around 20 seconds while using emulator, most of which from linear power spectrum computation, compared with about 80 seconds for the \texttt{MGCAMB} nonlinear calculation, saving much time in the process. This demonstrates that the emulator can be used as a fast diagnostic for MG nonlinear corrections, especially for identifying cases where linear observables alone are nearly degenerate.

\subsection{Emulator Accuracy in 3\texttimes 2pt Likelihood Predictions for DES-Y3}
\label{sec:theory-prediction}

For parameter constraints widely used in cosmology analyses, the emulator can be used to accelerate the likelihood calculation in MCMC analyses. For current observational likelihoods, we can use the emulator together with the linear power spectrum from \texttt{MGCAMB} to predict the nonlinear power spectrum in MG models, which is a key component in weak lensing likelihood 3$\times$2pt calculation. At present, some of the most precise weak lensing measurements are provided by DES Collaboration. It is important to know whether the emulator-induced bias is significant compared with current measurement uncertainties, and whether it is acceptable for the current weak lensing analyses. In this case, we show the emulator performance in the weak lensing likelihood theory prediction using DES-Y3 measurement errors as a reference of the possible bias.

\begin{figure}
    \centering
    \includegraphics[width=0.9\textwidth]{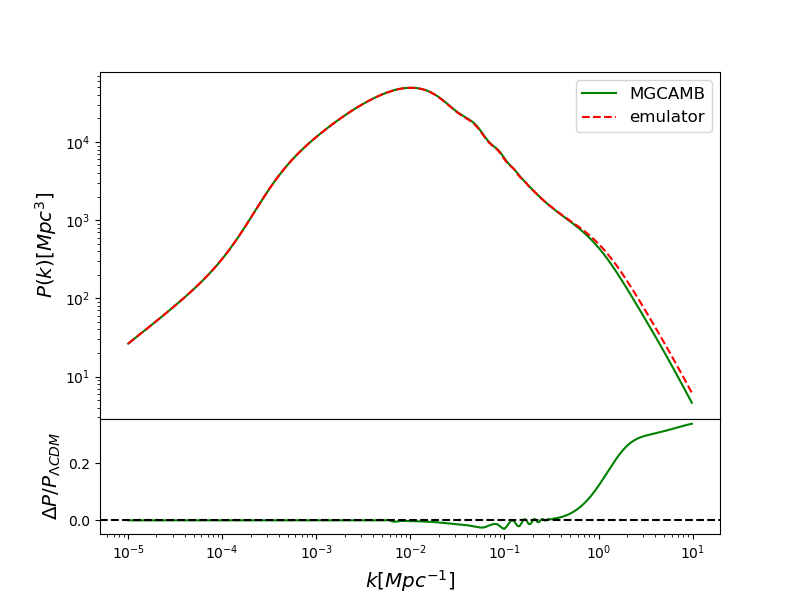}
    \caption{The nonlinear matter power spectrum for weak lensing calculation at $z=0$, with the emulator (red dashed line) and \texttt{MGCAMB} prediction (green solid line). The upper panel shows the nonlinear power spectra comparison and the lower panel shows the relative difference of power spectra between emulator and \texttt{MGCAMB}.}
    \label{fig:emulator-likelihood}
\end{figure}

\begin{figure}
    \centering
    \includegraphics[width=1.0\textwidth]{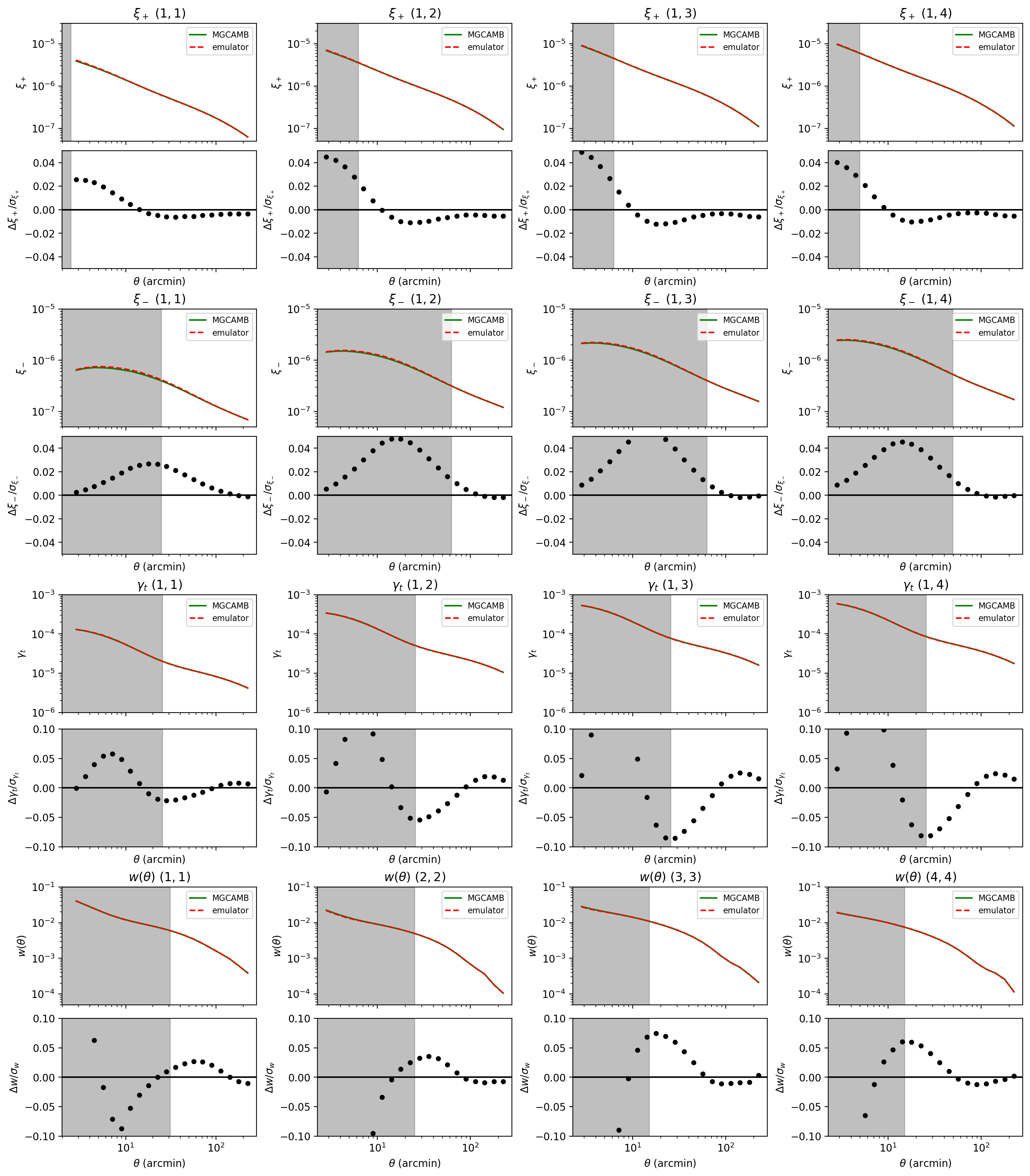}
    \caption{The weak lensing 3$\times$2pt theory prediction from the emulator (red dashed line) and \texttt{MGCAMB} (green solid line) in the worst case scenario. The upper panel shows their theory prediction comparison, and the lower panel shows the theory prediction differences divided by DES-Y3 errors. The grey region represents the scale-cut applied in DES-Y3 $\Lambda$CDM analysis.}
    \label{fig:emulator-3x2pt}
\end{figure}

Figure~\ref{fig:emulator-likelihood} shows the nonlinear matter power spectrum at $z=0$ chosen for the weak lensing calculation, with the emulator (red dashed line) and \texttt{MGCAMB} prediction (green solid line). The upper panel shows the nonlinear power spectra comparison and the lower panel shows the relative difference of power spectra between emulator and \texttt{MGCAMB}. This MG model is selected from the validation set used in Section~\ref{sec:validation-set} as one of the cases with the largest deviation in the nonlinear matter power spectrum between emulator and \texttt{MGCAMB} over all scales, considered as the worst-case scenario for the emulator performance. The relative difference between emulator and \texttt{MGCAMB} is about $10\%$ at $k=1~\mathrm{Mpc}^{-1}$ and reaches about $35\%$ at $k=10~\mathrm{Mpc}^{-1}$. In this worst case, the weak lensing 3$\times$2pt theory prediction from the emulator is shown in Figure~\ref{fig:emulator-3x2pt}. Here we plot all $w_{\theta}$ results in four redshift bins used in DES-Y3 results~\cite{DES:Y3LCDM}, while for $\xi_+$ and $\xi_-$ we only plot the first source bin with other four bins for clarity, and for $\gamma_t$ we only plot the first lens bin with other four source bins for clarity. In each plot of the correlation functions, the upper panel shows their theory prediction comparison with emulator shown as the red dashed line and \texttt{MGCAMB} as the green solid line, and the lower panel shows the theory prediction difference normalized by the DES-Y3 measurement uncertainty. The grey region represents the scale-cut applied in DES-Y3 $\Lambda$CDM analysis. We can see that even in this worst case, the relative difference of 3$\times$2pt theory prediction between emulator and \texttt{MGCAMB} is still smaller than the DES-Y3 measurement errors, with a maximum deviation of about $0.1\sigma$ in 3$\times$2pt theory prediction. This is because the scale-cut removes the small-scale observations, which is the region where the emulator has the largest deviation from \texttt{MGCAMB}. We also compute the $\chi^2$ difference in this MG model between the emulator and \texttt{MGCAMB} theory vectors, which gives  
\begin{equation}
    \Delta \chi^2=\boldsymbol d^{\,T}_{ref}C_{\mathrm{DES\text{-}Y3}}^{-1}\boldsymbol d_{ref}-\boldsymbol d^{\,T}_{emu}C_{\mathrm{DES\text{-}Y3}}^{-1}\boldsymbol d_{emu}=6.84,
\end{equation}
where $\boldsymbol d$ is the data vector after scale-cut, and $C_{\mathrm{DES\text{-}Y3}}$ is the covariance after point-mass correction. We emphasise that this is a deliberately selected extreme model and is not representative of the typical emulator behaviour. For a more typical case, e.g. MG model used in Section~\ref{sec:spectra-test} at $p_1=0$, we find $\Delta \chi^2=2.15$. This comparison quantifies the observable level discrepancy for the selected validation point, showing an acceptable level of agreement with 22 additional free parameters than standard cosmology, and therefore indicates that the emulator can be used to accelerate the likelihood calculation in MCMC analyses with current weak lensing measurements. We will present the results of the reconstruction using DES-Y3 and other cosmological datasets in a separate paper.

\subsection{Forecast for DESI + CSST}
\label{sec:forecast-desi-csst}

Using the two synthetic-data configurations introduced in Section~\ref{sec:mcmc}, we forecast the constraints achievable with the DESI-I+CSST-I and DESI-V+CSST-F survey combinations, and quantify the resulting improvements in the constraints on the reconstructed MG functions. We compare the constraints in a common mode basis to characterise how the constraining power changes between the two survey configurations.

As the reconstructed parameter nodes are strongly correlated, the constraining power is quantified more clearly in terms of principal components than by comparing individual nodes directly, providing an efficient way to summarize the constraining power~\cite{Zhao:2009fn,Hojjati:2011xd,Hall:2012te}. We perform a principal component analysis (PCA) of the reconstructed functions $\mu$, $\Sigma$ and $\Omega_X$ with all their sampled nodes.

To compare the two data combinations consistently, we define the PCA basis using the DESI-I+CSST-I covariance matrix and project the DESI-V+CSST-F covariance matrix onto the same basis. We choose the same number of principal components as nodes for both cases. The error of each principal component (eigenmode) is then given by the square root of the projected variances, which are diagonal elements of the rotated covariance matrix. These diagonal elements are eigenvalues for DESI-I+CSST-I, whereas the transformed DESI-V+CSST-F covariance need not be diagonal. The DESI-V+CSST-F mode amplitudes can therefore remain correlated in this common basis. In this method, we compare the two cases using the ratio of the PCA eigenmode errors defined as
\begin{equation}
    r_\alpha =
    \frac{\sigma_\alpha(\mathrm{DESI\mbox{-}V+CSST\mbox{-}F})}
         {\sigma_\alpha(\mathrm{DESI\mbox{-}I+CSST\mbox{-}I})},
\end{equation}
where $\alpha$ denotes the principal component and $r_\alpha<1$ indicates tighter constraints from the DESI-V+CSST-F data combination.

\begin{figure}[!htbp]
    \centering
    \includegraphics[width=0.88\textwidth]{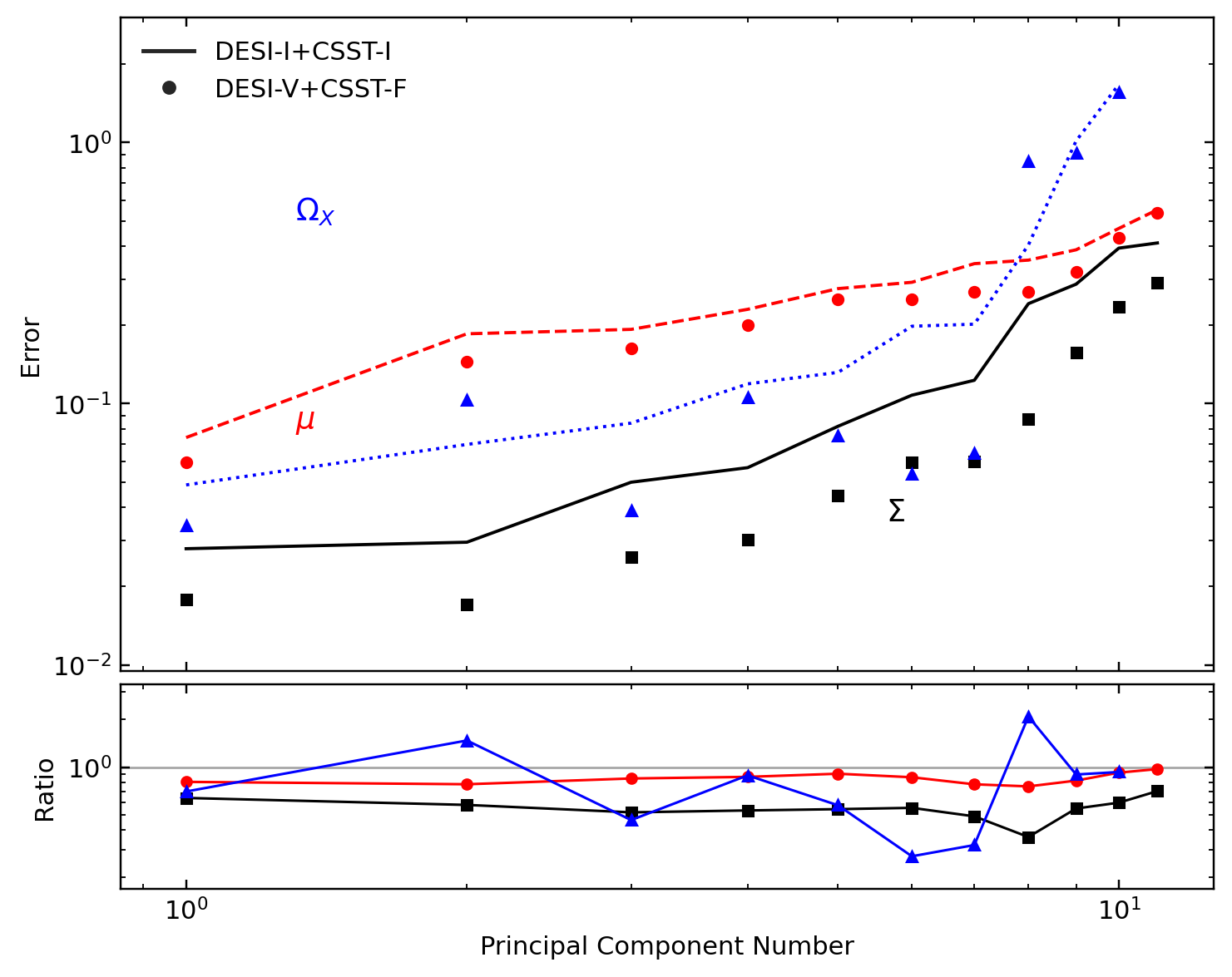}
    \caption{PCA eigenmode errors for the reconstructed $\Sigma$, $\mu$, and $\Omega_X$ functions. The upper panel shows the errors of the principal components for DESI-I+CSST-I and DESI-V+CSST-F. The lower panel shows the ratio $\sigma_{\mathrm{DESI\mbox{-}V+CSST\mbox{-}F}}/\sigma_{\mathrm{DESI\mbox{-}I+CSST\mbox{-}I}}$, where values below unity indicate tighter constraints from the DESI-V+CSST-F data combination. Lines represent the errors from DESI-I+CSST-I, while symbols represent those from DESI-V+CSST-F. $\mu$ errors are shown in red, $\Sigma$ in black, and $\Omega_X$ in blue.}
    \label{fig:pca-forecast}
\end{figure}

Figure~\ref{fig:pca-forecast} shows the PCA eigenmode errors for the reconstructed functions, with the corresponding numerical results given in Table~\ref{tab:pca-modes}. In the upper panel, the modes are ordered by increasing $\sigma_\alpha$ in the DESI-I+CSST-I reference basis. The lines show the eigenmode errors from DESI-I+CSST-I, while the symbols show those from DESI-V+CSST-F. The lower panel presents the ratio $r_\alpha$, for which values below unity indicate tighter constraints from DESI-V+CSST-F. Results for $\mu$, $\Sigma$, and $\Omega_X$ are shown in red, black, and blue, respectively. The numerical comparison in Table~\ref{tab:pca-modes} lists the PCA eigenmode errors for the three reconstructed functions separately, along with the ratio $r_\alpha$ and the dominant contributor to each eigenmode in the DESI-I+CSST-I basis.

The DESI-V+CSST-F combination generally reduces the eigenmode errors relative to DESI-I+CSST-I. For $\Sigma$, all 11 modes have $r_\alpha<1$, with a mean ratio of $0.551$, corresponding to an average error reduction of approximately $45\%$. The largest improvement occurs for the eighth $\Sigma$ mode, which is dominated by $\Sigma_3$ at $z=2$ and has $r_\alpha=0.361$. All 11 $\mu$ modes are also improved, with a mean ratio of $0.848$, corresponding to an average reduction of approximately $15\%$. The most improved $\mu$ mode is the eighth mode, dominated by $\mu_4$ at $z\simeq1.4$, with $r_\alpha=0.755$. For $\Omega_X$, 8 of the 10 modes have $r_\alpha<1$, with a mean ratio of $0.862$, corresponding to an average error reduction of approximately $14\%$. The largest improvement occurs for the sixth $\Omega_X$ mode, dominated by $\Omega_{X,9}$ at $z=0.2$, with $r_\alpha=0.273$. Overall, 30 of the 32 reconstructed PCA modes are more tightly constrained by DESI-V+CSST-F. The two $\Omega_X$ modes with $r_\alpha>1$ are the second mode dominated by $\Omega_{X,7}$ at $z=0.5$ and the eighth mode dominated by $\Omega_{X,3}$ at $z=2$, with $r_\alpha=1.475$ and $2.093$, respectively. Their broader projected constraints may be associated with differences in the redshift sampling of the two DESI-like synthetic datasets. In DESI-V, the reduced number of intermediate-redshift measurements between $z=0.2$ and $z=0.75$, together with the absence of the Ly$\alpha$ measurement at $z=2.33$, reduces the constraining power around these nodes. The PCA forecast therefore indicates that the higher-precision DESI-V+CSST-F combination improves the majority of the reconstruction modes defined by the DESI-I+CSST-I covariance, while some $\Omega_X$ modes remain sensitive to the number and redshift coverage of the DESI-V sampling points. The smaller average improvement in $\mu$ than in $\Sigma$ may also reflect this limitation.

\begin{table}[!htbp]
    \centering
    \tiny
    \setlength{\tabcolsep}{2pt}
    \resizebox{\textwidth}{!}{%
    \begin{tabular}{@{}ccclccclcccl@{}}
        \toprule
        \multicolumn{4}{c}{$\Sigma$ PCA modes} &
        \multicolumn{4}{c}{$\mu$ PCA modes} &
        \multicolumn{4}{c}{$\Omega_X$ PCA modes} \\
        \cmidrule(lr){1-4}\cmidrule(lr){5-8}\cmidrule(lr){9-12}
	        Mode & $\sigma_{\rm I}/\sigma_{\rm V}$ & $r_\alpha$ & Dominant &
	        Mode & $\sigma_{\rm I}/\sigma_{\rm V}$ & $r_\alpha$ & Dominant &
	        Mode & $\sigma_{\rm I}/\sigma_{\rm V}$ & $r_\alpha$ & Dominant \\
        \midrule
	         1 & 0.0278/0.0178 & 0.638 & $\Sigma_6(+0.839)$ &
	         1 & 0.0742/0.0598 & 0.807 & $\mu_1(+0.966)$ &
	         1 & 0.0488/0.0342 & 0.701 & $\Omega_{X,10}(+0.536)$ \\
	         2 & 0.0295/0.0170 & 0.577 & $\Sigma_5(+0.673)$ &
	         2 & 0.185/0.144 & 0.779 & $\mu_2(+0.850)$ &
	         2 & 0.0698/0.103 & 1.475 & $\Omega_{X,7}(+0.839)$ \\
	         3 & 0.0500/0.0259 & 0.517 & $\Sigma_7(+0.653)$ &
	         3 & 0.192/0.163 & 0.848 & $\mu_7(+0.562)$ &
	         3 & 0.0842/0.0389 & 0.462 & $\Omega_{X,6}(+0.912)$ \\
	         4 & 0.0569/0.0303 & 0.532 & $\Sigma_4(+0.695)$ &
	         4 & 0.230/0.199 & 0.868 & $\mu_8(+0.556)$ &
	         4 & 0.119/0.106 & 0.887 & $\Omega_{X,5}(+0.910)$ \\
	         5 & 0.0819/0.0444 & 0.542 & $\Sigma_9(+0.539)$ &
	         5 & 0.276/0.250 & 0.909 & $\mu_7(+0.523)$ &
	         5 & 0.132/0.0755 & 0.573 & $\Omega_{X,8}(+0.757)$ \\
	         6 & 0.108/0.0595 & 0.552 & $\Sigma_4(+0.554)$ &
	         6 & 0.291/0.251 & 0.863 & $\mu_3(+0.574)$ &
	         6 & 0.198/0.0539 & 0.273 & $\Omega_{X,9}(+0.737)$ \\
	         7 & 0.123/0.0600 & 0.488 & $\Sigma_{10}(+0.843)$ &
	         7 & 0.343/0.267 & 0.779 & $\mu_5(+0.630)$ &
	         7 & 0.201/0.0647 & 0.321 & $\Omega_{X,4}(+0.890)$ \\
	         8 & 0.241/0.0871 & 0.361 & $\Sigma_3(+0.472)$ &
	         8 & 0.354/0.268 & 0.755 & $\mu_4(+0.702)$ &
	         8 & 0.405/0.847 & 2.093 & $\Omega_{X,3}(+0.834)$ \\
	         9 & 0.287/0.157 & 0.547 & $\Sigma_3(+0.718)$ &
	         9 & 0.388/0.319 & 0.822 & $\mu_{10}(+0.613)$ &
	         9 & 1.01/0.910 & 0.898 & $\Omega_{X,2}(+0.785)$ \\
	        10 & 0.394/0.235 & 0.596 & $\Sigma_2(+0.755)$ &
	        10 & 0.468/0.432 & 0.923 & $\mu_{11}(+0.640)$ &
	        10 & 1.67/1.55 & 0.932 & $\Omega_{X,1}(+0.949)$ \\
	        11 & 0.412/0.290 & 0.705 & $\Sigma_{11}(+0.981)$ &
	        11 & 0.552/0.537 & 0.972 & $\mu_{11}(+0.605)$ &
         & & & \\
        \bottomrule
    \end{tabular}}
    \caption{PCA comparison between DESI-I+CSST-I and DESI-V+CSST-F for the reconstructed $\Sigma$, $\mu$, and $\Omega_X$ functions. The PCA basis is defined by the DESI-I+CSST-I covariance. The $\sigma_{\rm I}/\sigma_{\rm V}$ column lists the DESI-I+CSST-I and DESI-V+CSST-F eigenmode errors, respectively. The ratio is defined as $r_\alpha=\sigma_{\mathrm{V+F}}/\sigma_{\mathrm{I+I}}$, so values below unity indicate tighter projected constraints in DESI-V+CSST-F. The dominant column lists the largest absolute contributor to each eigenmode in the DESI-I+CSST-I basis, with the corresponding coefficient in parentheses.}
    \label{tab:pca-modes}
\end{table}

\section{Discussion}
\label{sec:discussion}

In this work, we have constructed an emulator with a neural network method for the matter power spectrum MG nonlinear correction in the model-independent reconstruction. Instead of emulating the full nonlinear spectrum directly, the neural network predicts the ratio $R_{\mathrm{MG}}=P_{\mathrm{MG}}^{\mathrm{NL}}/P_{\mathrm{MG}}^{\mathrm{L}}$ in the range $10^{-5} \leq k < 10\,\mathrm{Mpc}^{-1}$. This keeps the linear theory prediction from \texttt{MGCAMB} and isolates the nonlinear part of the calculation, which is expensive in repeated likelihood evaluations. The emulator is trained on a total of 209772 spectra, using both samples from previous reconstruction chains and randomly generated points covering the adopted $3\sigma$ parameter posteriors. The input training parameters include the standard cosmological parameters, the reconstructed $\mu(a)$ and $\Omega_X(a)$ nodes, redshift, and the nonlinear parameter $p_1$.

We performed several validation tests to assess the emulator performance. In the representative spectra comparison, the emulator reproduces the reference nonlinear power spectra to within $1.5\%$ for $\Lambda$CDM and for the selected MG models with $p_1=0$ and $p_1=1$. For the stronger nonlinear MG case with $p_1=2$, the same level of agreement is maintained for $k<1\,\mathrm{Mpc}^{-1}$, while larger deviations appear at higher $k$, where the nonlinear MG correction becomes stronger. The 2000 independent validation points provide a broader statistical check: the mean residual is close to zero, the $2\sigma$ scatter stays below $1\%$ for $k<0.5\,\mathrm{Mpc}^{-1}$, and it increases to roughly $2\%$ on smaller scales.  An additional test of 2000 out-of-range MCMC points finds a mean residual close to zero but broader small-scale scatter, with the $2\sigma$ region extending up to approximately $4\%$. These tests indicate that the significant emulator error mainly occurs in the strongly nonlinear, high-$k$ regime rather than as a systematic error across all the scales and the validation set.

The synthetic data MCMC tests give a parameter-constraining validation. Using DESI-like BAO+RSD data and CSST-like angular power spectrum data generated from a known $\Lambda$CDM fiducial model, the result recovers the input cosmological parameters within the posterior uncertainties. The reconstructed $\Omega_X(a)$ function is also consistent with the fiducial background history. The best-fit $\mu(a)$ and $\Sigma(a)$ functions are generally close to their GR limits, with the largest deviations appearing at very low redshift where the data have limited constraining power. In the DESI-I+CSST-I case, the posterior-mean shifts in $\mu$ and $\Sigma$ remain within the $1\sigma$ uncertainties and can be mainly attributed to projection effects with the galaxy bias parameters. These shifts are reduced in the higher-precision DESI-V+CSST-F case, supporting the conjecture that their deviations are driven by projection effects and data precision rather than by the emulator errors.

In practice, the emulator is useful as a diagnostic for nonlinear information from MG. The MG reconstruction model example is the same as that used in the validation tests. Since the emulator predicts $R_{\mathrm{MG}}$ directly, it can expose these nonlinear differences without recomputing the full nonlinear power spectrum. This provides a practical way to identify MG signals that are hidden under the linear perturbation degeneracy and to assess how much the nonlinear information can help distinguish structure formation histories in different gravity theories.

We also show the possibility of another application of the emulator in weak lensing likelihood theory predictions, using DES-Y3 $3\times2$pt measurements as an example and to quantify the impact of emulator systematic errors. We selected a point from the validation set with the largest deviations from \texttt{MGCAMB}, for which the emulator error on power spectrum reaches about $10\%$ at $k=1\,\mathrm{Mpc}^{-1}$ and about $35\%$ at $k=10\,\mathrm{Mpc}^{-1}$. After being projected into the $3\times2$pt correlation functions and applying DES-Y3 scale cuts, the theory vector difference remains below the DES-Y3 measurement uncertainties, with a maximum deviation of about $0.1\sigma$, while the $\chi^2$ difference between emulator and \texttt{MGCAMB} shows $\Delta \chi^2=6.84$ with 22 additional free parameters than standard cosmology in this deliberately selected extreme model. This does not replace a real data analysis, but it shows that the emulator error is subdominant compared to the current weak-lensing precision, and the emulator is accurate enough for parameter constraints of current photometric surveys.

Finally, we use the emulator to forecast the constraining power of DESI+CSST synthetic data on the reconstructed functions. Since the parameter nodes are highly correlated, we perform PCA of $\Sigma$, $\mu$, and $\Omega_X$ to quantify the forecast improvement. Using the basis defined by the DESI-I+CSST-I covariance, we project the DESI-V+CSST-F covariance onto the same modes and compare their projected errors. The higher-precision DESI-V+CSST-F case improves all $\Sigma$ and $\mu$ modes and 8 of the 10 $\Omega_X$ modes, corresponding to 30 of the 32 modes in total. The average error reductions are approximately $45\%$, $15\%$, and $14\%$ for $\Sigma$, $\mu$, and $\Omega_X$, respectively. The largest reduction for an individual mode occurs in the $\Omega_X$ sector, while the only two modes that become less constrained are also associated with $\Omega_X$. This suggests that the background reconstruction is more sensitive to survey redshift sampling and degeneracy directions than the $\Sigma$ and $\mu$ reconstructions. These PCA results demonstrate how future galaxy surveys can improve model-independent MG reconstruction when the emulator makes the required nonlinear calculations computationally feasible.

Several limitations should be noted. The emulator should only be used within the training $k$ range, extrapolation beyond the range is not controlled. We need to be careful when using the emulator outside training parameter space as the prediction accuracy may be worsened. The reconstruction here is assumed scale independent, thus only the nonlinear parameter $p_1$ is varied while the other nonlinear parameters in \texttt{ReACT} are fixed. Emulator accuracy degrades for large $p_1$ and high $k$, suggesting that it should be used with caution in extremely MG nonlinear regimes. The current \texttt{ReACT} calculation is calibrated up to $z=2.5$, while the redshift bins of CSST and other current photometric surveys (e.g. DES) extend to a higher redshift. We therefore treat the contribution from the high-redshift regime as a controlled approximation as is implemented in \texttt{MGCAMB}. Finally, emulator systematic errors do not significantly affect the posterior of parameters, while the tests are based on synthetic data rather than analyses of observed data, and real data are only used to demonstrate the applicability of the emulator. Future work should extend the parameter training range, include additional nonlinear degrees of freedom, extend to other MG models, and apply the emulator to parameter constraints of current photometric surveys.

\section*{Acknowledgements}

We thank Alessio Spurio Mancini for helpful discussions on the use of \texttt{CosmoPower} and emulator training, and Levon Pogosian for providing the reconstruction chains used in constructing the emulator training set. LY and GBZ are supported by the National Key R \& D Program of China (2023YFA1607800, 2023YFA1607803), NSFC grant 12525301, and by the CAS Project for Young Scientists in Basic Research (No. YSBR-092). LY is also supported by the Chinese Scholarship Council (CSC) and the University of Portsmouth. GBZ is also supported by science research grants from the China Manned Space Project with No. CMS-CSST-2021-B01, and the New Cornerstone Science Foundation through the XPLORER prize. KK is supported also by STFC grant number ST/B001175/1. This work is also supported by science research grants from the China Manned Space Project with grant Nos. CMS-CSST-2025-A02.

Numerical computations were carried out on the \textsc{Sciama} High Performance Computing (HPC) cluster, which is supported by the Institute of Cosmology and Gravitation, the South-East Physics Network (SEPNet) and the University of Portsmouth. 

For the purpose of open access, we have applied a Creative Commons Attribution (CC BY) licence to any Author Accepted Manuscript version arising. Supporting research data are available on reasonable request. 

\bibliographystyle{JHEP}
\bibliography{bib/references}

\end{document}